\documentclass[journal, twoside]{IEEEtran}

\usepackage{cite}
\usepackage{amsmath,amssymb,amsfonts}
\usepackage{algorithm}
\usepackage{algpseudocode}
\usepackage{graphicx}
\usepackage{xcolor}
\usepackage{xspace}
\usepackage{booktabs}
\usepackage{pifont}
\usepackage{colortbl}
\usepackage[group-separator={,}, detect-weight]{siunitx}
\usepackage{tikz}
\usepackage{subcaption}
\usepackage{makecell}
\usepackage{orcidlink}
\usepackage{multirow}
\usepackage{hyperref}
\def\BibTeX{{\rm B\kern-.05em{\sc i\kern-.025em b}\kern-.08em
    T\kern-.1667em\lower.7ex\hbox{E}\kern-.125emX}}

\newcommand{\ie}{{\it i.e.,}\xspace}

\newcommand{\sysname}{{\it SwiftChannel}\xspace}

\newcommand{\colorbarClock}[2]{%
\begin{tikzpicture}
        \fill[red!30] (0,0) rectangle (#1*2,0.3);
    \node[anchor=center, text width=1.8cm, align=center] at (1.1,0.15) {#2};
\end{tikzpicture}
}
\newcommand{\colorbarBRAM}[2]{%
\begin{tikzpicture}
    \fill[red!30] (0,0) rectangle (#1*2,0.3);
    \node[anchor=center, text width=1.8cm, align=center] at (1.1,0.15) {#2};
\end{tikzpicture}
}
\newcommand{\colorbarDSP}[2]{%
\begin{tikzpicture}
    \fill[red!30] (0,0) rectangle (#1*2,0.3);
    \node[anchor=center, text width=1.8cm, align=center] at (1.1,0.15) {#2};
\end{tikzpicture}
}
\newcommand{\colorbarFF}[2]{%
\begin{tikzpicture}
    \fill[red!30] (0,0) rectangle (#1*2,0.3);
    \node[anchor=center, text width=1.8cm, align=center] at (1.1,0.15) {#2};
\end{tikzpicture}
}
\newcommand{\colorbarLUT}[2]{%
\begin{tikzpicture}
    \fill[red!30] (0,0) rectangle (#1*2,0.3);
    \node[anchor=center, text width=1.8cm, align=center] at (1.1,0.15) {#2};
\end{tikzpicture}
}

\begin{document}

\title{\sysname: Algorithm-Hardware Co-Design for Deep Learning-Based 5G Channel Estimation}

\author{
Shengzhe~Lyu~\orcidlink{0009-0004-7331-7700},~\IEEEmembership{Student~Member,~IEEE,}
Yuhan~She~\orcidlink{0000-0003-3748-577X},~\IEEEmembership{Student~Member,~IEEE,}
Di~Duan~\orcidlink{0000-0003-4184-0762},~\IEEEmembership{Member,~IEEE,}
Tao~Ni~\orcidlink{0000-0003-0671-3020},~\IEEEmembership{Member,~IEEE,}
Yu~Hin~Chan~\orcidlink{0009-0005-8309-9898},
Chengwen~Luo~\orcidlink{0000-0003-0293-0781},~\IEEEmembership{Member,~IEEE,}\\
Ray~C.~C.~Cheung\textsuperscript{\textasteriskcentered}~\orcidlink{0000-0002-6764-0729},~\IEEEmembership{Senior Member,~IEEE,}
and~Weitao~Xu\textsuperscript{\textasteriskcentered}~\orcidlink{0000-0001-9741-5912},~\IEEEmembership{Senior Member,~IEEE}
% \thanks{Manuscript received April 19, 2005; revised April 19, 2005; accecpted April 19, 2005. Date of publication 19 August 2005; date of current version 19 August 2005. This work was supported in part by TBD. Recommended for acceptance by TBD. \textit{(Corresponding author: TBD.)}}
% \thanks{\vspace{30pt}}
\thanks{Shengzhe Lyu and Weitao Xu are with the Department of Computer Science, City University of Hong Kong, Hong Kong (e-mail: shengzhe.lyu@my.cityu.edu.hk; weitaoxu@cityu.edu.hk).}
\thanks{Yuhan She, Yu Hin Chan, and Ray C. C. Cheung are with the Department of Electrical Engineering, City University of Hong Kong, Hong Kong (e-mail: yuhanshe3-c@my.cityu.edu.hk; yhchan96@cityu.edu.hk; r.cheung@cityu.edu.hk).}
\thanks{Di Duan is with the Department of Information Engineering, The Chinese University of Hong Kong, Hong Kong (e-mail: duandiacademic@gmail.com).}
\thanks{Tao Ni is with the Computer, Electrical and Mathematical Sciences and Engineering Division, King Abdullah University of Science and Technology, Saudi Arabia (e-mail: tao.ni@kaust.edu.sa).}
\thanks{Chengwen Luo is with the National Engineering Laboratory for Big Data System Computing Technology, Shenzhen University, Shenzhen 518060, China (e-mail: chengwen@szu.edu.cn).}
% \thanks{Digital Object Identifier 10.1109/TMC.XXXX.XXXXXXX}
\thanks{* Weitao~Xu and Ray~C.~C.~Cheung are the corresponding authors.}
}

\markboth{IEEE TRANSACTIONS ON MOBILE COMPUTING,~Vol.~14, No.~8, August~2015}%
{LYU \MakeLowercase{\textit{et al.}}: SWIFTCHANNEL: ALGORITHM-HARDWARE CO-DESIGN FOR DEEP LEARNING-BASED 5G CHANNEL ESTIMATION}

\maketitle

\begin{abstract}
Channel estimation is crucial in 5G communication networks for optimizing transmission parameters and ensuring reliable, high-speed communication.
However, the use of multiple-input and multiple-output (MIMO) and millimeter-wave (mmWave) in 5G networks presents challenges in achieving accurate estimation under strict latency requirements on resource-limited hardware platforms.
To address these challenges, we propose \textit{SwiftChannel}, an algorithm-hardware co-design framework that integrates a hardware-friendly deep learning-based channel estimator with a dedicated accelerator.
Our approach employs a convolutional neural network enhanced with a parameter-free attention mechanism, which effectively reconstructs full-resolution spatial-frequency domain channel matrices from low-resolution least squares (LS) estimates.
We further develop a multi-stage model compression pipeline combining knowledge distillation, convolution re-parameterization, and quantization-aware training, resulting in substantial model size reduction with negligible accuracy loss.
The hardware accelerator, implementing the compressed model and the LS estimator on FPGA platforms using High-level Synthesis (HLS), features a fine-grained pipeline architecture and optimized dataflow strategies. Tested on a Zynq UltraScale+ RFSoC, the accelerator achieves sub-millisecond latency, providing up to 24x speed-up and over 33x improvement in energy efficiency compared to GPU-based solutions.
Extensive evaluations demonstrate that the proposed design generalizes not only across various noise levels and user mobilities, but also to a variety of unseen channel profiles, outperforming
state-of-the-art baselines.
By unifying algorithmic innovation with hardware-aware design, our work presents a future-proof channel estimation solution for 5G MIMO systems.
The source codes for the dataset synthesis, deep learning algorithm, and HLS-based FPGA design are accessible via GitHub\footnote{\url{https://github.com/shengzhelyu65/SwiftChannel}}.
\end{abstract}


\IEEEpeerreviewmaketitle

\section{Introduction}
\subsection{Background and Motivation}
\IEEEPARstart{C}{hannel} estimation serves as the cornerstone of 5G and future wireless communication systems, enabling reliable data transmission, spectral efficiency, and massive connectivity.
Its importance is amplified by the adoption of massive multiple-input and multiple-output (MIMO) and millimeter-wave (mmWave) technologies, which introduce unprecedented spatial and spectral dimensions.
In the spatial domain, massive MIMO systems with dozens to hundreds of antennas require precise channel state information to dynamically steer narrow beams toward users, where estimation inaccuracies directly degrade beamforming gain and amplify inter-user interference.
In the frequency domain, mmWave bands suffer from severe path loss, blockage sensitivity, and rapid fluctuations, necessitating real-time tracking of variations in dense urban environments.

\begin{figure}[t]
  \centering
  \includegraphics[width=\linewidth]{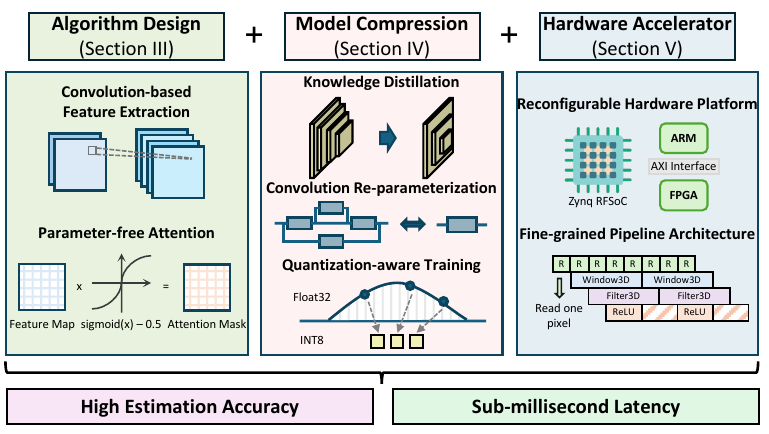}
  \caption{The proposed \sysname achieves high-accuracy and low-latency channel estimation by integrating algorithm design, model compression, and hardware accelerator design.}
  \label{fig:teaser_figure}
  \vspace{-15pt}
\end{figure}

Although traditional channel estimation methods based on statistical techniques such as least squares (LS) and linear minimum mean square error (LMMSE)~\cite{van1995channel, li1998robust, alwazani2020intelligent} are computationally efficient and widely used in modern base stations, they fall short in accuracy.
LS suffers from noise sensitivity, while LMMSE relies on impractical assumptions about prior channel statistics.
With the recent advancements in deep learning (DL) techniques, attention has shifted toward leveraging neural networks to enhance estimation performance following LS or LMMSE results, as those methods could exploit their powerful representation learning capabilities to model the complex, nonlinear relationships inherent in wireless channels~\cite{soltani2019deep, li2019deep, luan2021low, luan2023channelformer, yang2021deep, zhou2023pay, zhou2025low, sharma2024low}.
However, most DL-based methods focus solely on algorithmic innovation without addressing hardware feasibility, especially for resource-constrained hardware, hindering their practical applicability in real-world scenarios.
As such, \textit{how to achieve high-accuracy real-time channel estimation in 5G MIMO systems} remains an open and challenging research question.

To bridge the gap, algorithm-hardware co-design is emerging as a promising solution that integrates hardware-aware algorithm development with customized hardware accelerator design to optimize performance.
The neural network is deliberately designed to accommodate the constraints of the target hardware platform, such as memory bandwidth, DSP slices, and latency requirements.
This is achieved by adopting hardware-friendly architectures and incorporating model compression techniques to ensure smooth deployment.
In turn, the hardware architecture is designed to enhance the algorithm's parallelism during model inference by utilizing specialized dataflow strategies and optimized memory access patterns.
This bidirectional optimization ensures an effective balance between computational performance and hardware utilization.
In line with this approach, recent works like~\cite{sharma2024low, haq2023deep, haq2024low, chundi2021channel, mirfarshbafan2020beamspace} explore the implementation of DL-aided wireless channel estimation algorithms on field-programmable gate arrays (FPGAs) or application-specific integrated circuits (ASICs) for high energy efficiency and low processing delay.
However, most existing works compromise on model complexity in order to meet hardware constraints, resulting in reduced estimation accuracy.
Furthermore, few studies investigate the functionality and effectiveness of model compression methods prior to implementation, leaving this field underexplored.

\subsection{Challenges and Contributions}
In this paper, we introduce \sysname, a novel algorithm-hardware co-design, as illustrated in Fig.~\ref{fig:teaser_figure}. Our method involves training a deep learning-based algorithm, applying model compression, and subsequently deploying it on a reconfigurable platform. This approach effectively balances model complexity with hardware constraints, optimizing both the performance of the channel estimation and the feasibility of its practical implementation. Furthermore, it addresses the following three critical challenges:

\textbf{(1) Input Sparsity in Spatial and Frequency Domains.}
In 5G massive MIMO systems employing hybrid beamforming, the precoding structure introduces input sparsity in both spatial and spectral domains.
To optimize power efficiency and reduce antenna switching times, only a limited subset of base station antennas is utilized for receiving pilot signals, limited by the total number of radio frequency (RF) chains~\cite{jiang2022accurate, zhou2025low, gao2023deep}.
Similarly, in the frequency domain, user equipment (UE) transmits pilot signals on selected subcarriers determined by the transmission comb type, resulting in a compressed spectral domain representation of the received signal~\cite{3gpp.38.211}.
Consequently, the dimensions of the LS-estimated channel matrix, which serves as the input to the proposed neural network, are significantly smaller than the full channel matrix.
This presents a two-fold challenge for DL-based algorithm design: it must effectively extract critical channel features from these sparse LS estimations while simultaneously upscaling the low-resolution inputs to reconstruct the complete spatial-frequency domain channel matrix accurately.

\textbf{(2) Sub-Millisecond Latency Constraint.}
In the 3GPP 5G New Radio (NR) specifications, the Sounding Reference Signal (SRS) serves as the pilot signal for channel estimation. As a result, the processing latency of each received SRS symbol should be aligned with the SRS transmission interval to avoid outdated estimated channel state information.
Considering orthogonal frequency-division multiplexing (OFDM) modulation, the subcarrier spacing is \(120\,\mathrm{kHz}\) for a numerology of \(3\), resulting in a slot duration of \(0.125\,\mathrm{ms}\). With SRS transmitted every \(8\) slots, the total interval is \(1\,\mathrm{ms}\)~\cite{3gpp.38.211}.
Consequently, both the algorithm and its hardware implementation must be optimized to operate within sub-millisecond latency to ensure the result remains valid.
However, this constraint severely limits algorithmic complexity, precluding the use of computationally intensive structures like self-attention modules proposed in prior works~\cite{luan2023channelformer, yang2021deep, zhou2023pay, zhou2025low}.
Even with simpler architectures, such as those employed in algorithm-hardware co-design approaches~\cite{sharma2024low, haq2023deep, haq2024low}, it is still challenging to achieve the required latency demands without model compression before implementation.

\textbf{(3) Power Limitation and Hardware Compatibility.}
The hardware platform that implements the proposed algorithm must meet not only the performance requirements but also the power limitations of current 5G infrastructures, which have a low power budget.
It should also ensure seamless compatibility with other components, such as RF front-ends and baseband processors.
Additionally, the design should include sufficient flexibility to accommodate future upgrades and evolving standards.
While GPUs are generally efficient for deep learning model inference, they usually consume more energy due to high operating clock frequencies.
In contrast, customized architectures, particularly reconfigurable designs on FPGAs, operate at lower frequencies and provide specialized solutions for improved energy efficiency.
However, these platforms are inherently less general and flexible compared to GPUs.
Consequently, a key challenge lies in selecting the most suitable hardware platform and devising a design strategy that balances performance, efficiency, and adaptability while facilitating easy upgrades in the future.

We propose several novel techniques to address the challenges mentioned above. \textbf{\underline{(1)}} Specifically, to tackle input sparsity, we conceptualize the LS-estimated channel matrix as a low-resolution image and the full channel matrix as its high-resolution counterpart. Drawing inspiration from image super-resolution, the proposed neural network is designed to extract critical features from sparse inputs and upscale them into full dimension.
This design incorporates a parameter-free attention mechanism, which is a lightweight technique that allows the model to selectively focus on and refine important channel features, and employs pixel shuffle for efficient upscaling, avoiding additional computational overhead.
\textbf{\underline{(2)}} To meet latency constraints, we further compress the model through knowledge distillation, re-parameterization, and quantization, which reduces the model size and parameter count.
The compression algorithm is implemented targeting reconfigurable hardware platforms featuring powerful FPGAs that maximize parallelism and minimize processing delay.
\textbf{\underline{(3)}} FPGAs typically offer significantly higher power efficiency compared to GPUs or CPUs. To improve the design's generalizability, we utilize High-Level Synthesis (HLS) to translate high-level abstractions into register-transfer level designs, ensuring its adaptability across various SoC platforms and making the hardware implementation platform-independent. Additionally, the Advanced eXtensible Interface (AXI) protocol is employed to enhance interoperability, facilitating seamless integration with other hardware components for future upgrades.

To the best of our knowledge, this work represents the first systematic algorithm-hardware co-design for 5G MIMO channel estimation, seamlessly integrating neural network architecture, model compression, and hardware implementation. Our primary contributions are:
\begin{itemize}
    \item We propose a novel deep learning-based algorithm for 5G MIMO channel estimation that reconstructs the full channel matrix by scaling LS estimations across both frequency and spatial domains. The model incorporates a parameter-free attention mechanism, which allows it to selectively focus on regions with richer information, thereby enhancing estimation accuracy while avoiding unnecessary computational overhead. (See Section~\ref{sec:network})
    \item We present a three-stage model compression framework designed to reduce model size while maintaining high accuracy, resulting in a hardware-aware compressed network. Knowledge distillation is first utilized to transfer the knowledge to a smaller student model. Then the post-training re-parameterization reduces the model size by restructuring the convolutional layers. Finally, we perform quantization-aware training to enable efficient integer-based computations. (See Section~\ref{sec:optimization})
    \item We design a customized hardware accelerator using HLS, implementing both the LS estimator and a compressed neural network on FPGAs. The design can be adapted for various platforms and is employed on a Zynq UltraScale+ RFSoC for testing. Featuring a fine-grained pipeline architecture and on-chip weight caches, the accelerator provides real-time performance with sub-millisecond latency and considerable energy efficiency improvements compared to GPU-based solutions. (See Section~\ref{sec:hardware})
\end{itemize}

The rest of the paper is organized as follows.
Section~\ref{sec:related_work} reviews related works.
Section~\ref{sec:system_model} outlines the channel estimation process and the system model.
Sections~\ref{sec:network},~\ref{sec:optimization}, and~\ref{sec:hardware} present the network design, the model compression techniques, and the hardware accelerator design, respectively.
Section~\ref{sec:evaluation} provides the evaluation of the co-design.
Finally, Section~\ref{sec:conclusion} summarizes the key findings of the paper.

\section{Related Work}
\label{sec:related_work}  
\subsection{Algorithmic Channel Estimation Approaches}
Classical channel estimation techniques, including LS~\cite{van1995channel} and LMMSE~\cite{li1998robust, alwazani2020intelligent, pourkabirian2021robust}, rely on pilot signals and statistical models. While computationally efficient, these methods face challenges in high-mobility or millimeter-wave scenarios due to rapid channel variations and sparse scattering environments. Compressed sensing (CS)-based approaches~\cite{lin2022channel, chen2023channel, gao2015spatially} exploit channel sparsity in specific domains for full-channel recovery, though their reliance on iterative solvers often increases computational latency.

Deep learning enables data-driven channel estimation by leveraging powerful data-fitting and feature extraction capabilities, enabling the learning of intrinsic channel properties from extensive training data.
Convolutional Neural Networks (CNNs) excel in this domain by exploiting feature extraction to analyze channel frequency and temporal characteristics.
Building on classical CNNs, Dong et al.~\cite{dong2019deep} introduced spatial-frequency CNN and spatial-frequency-time CNN architectures for millimeter-wave massive MIMO systems. Jiang et al.~\cite{jiang2021dual} addressed high noise interference and pilot contamination through a dual-CNN architecture.
Soltani et al.~\cite{soltani2019deep} integrated a two-stage framework combining SRCNN and DnCNN for super-resolution and denoising, while Li et al.~\cite{li2019deep} employed residual networks to improve estimation performance, utilizing transposed convolution for upsampling.
To reduce the complexity introduced by the upscaling layer, Luan et al.~\cite{luan2021low} investigated replacing the layer with bilinear interpolation.
While these works demonstrate the effectiveness of various CNN backbones, our work excels by integrating a parameter-free attention mechanism for feature refinement with a holistic, multi-stage compression framework.

Beyond conventional paradigms, Jiang et al.~\cite{jiang2021learning} utilized Graph Neural Networks (GNNs) to directly map pilot signals to system configurations.
Attention mechanisms, which have achieved significant success in fields like natural language processing, have inspired models such as Channelformer~\cite{luan2023channelformer} and AttenFreqTimeNet~\cite{yang2021deep}, demonstrating improved performance through long-range dependency capture in time-frequency domains.
Zhou et al.~\cite{zhou2025low} proposed KDD-SFCEN, a 3D attention-based framework for spatial-frequency-temporal channel extrapolation, significantly enhancing accuracy with low pilot overhead.
These methods highlight the efficacy of attention mechanisms but do not analyze the feasibility of deploying computationally intensive implementations on hardware.
Apart from these data-driven models, model-driven frameworks such as~\cite{gao2023deep, gao2018comnet} integrate physical channel models with neural networks.

\subsection{Channel Estimation Algorithm-Hardware Co-Designs}
Hardware-friendly algorithm design and dedicated hardware acceleration co-design for channel estimation remain underexplored, with key challenges lying in balancing latency, resource utilization, and numerical precision.
Sharma et al.~\cite{sharma2024low} developed a co-design framework for LS-augmented neural networks with dense layers on a Zynq SoC, achieving reduced latency through hardware-software partitioning.
Haq et al.~\cite{haq2023deep} implemented a classical model architecture with dense layers for preamble-based OFDM system channel estimation on an SoC and further evaluated its performance, power, and area on a 45-nm ASIC.
In~\cite{haq2024low}, separate networks were designed for the real and imaginary components of received pilot signals prior to FPGA implementation, demonstrating enhanced estimation accuracy.
Chundi et al.~\cite{chundi2021channel} leveraged sparsity in a model-based deep learning approach, reducing overall energy consumption on FPGA compared to CPU and GPU.
Mirfarshbafan et al.~\cite{mirfarshbafan2020beamspace} proposed an adaptive denoising method for beamspace domain channel vectors, accompanied by FPGA implementation results validating its efficacy.
In contrast, our work presents a fine-grained streaming accelerator, addressing end-to-end dataflow challenges from feature extraction to the complex data reorganization required by pixel shuffling.

\section{System Model}
\label{sec:system_model}
\begin{figure}[t]
  \centering
  \includegraphics[width=0.9\linewidth]{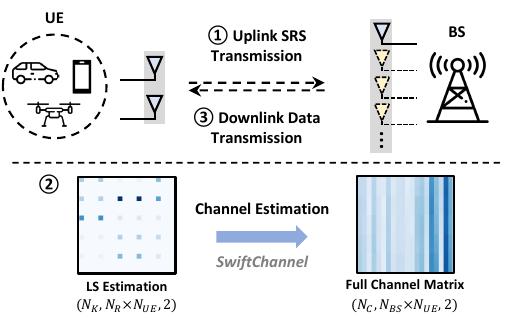}
  \caption{The uplink channel estimation process.}
  \label{fig:background_channel_estimation}
  \vspace{-8pt}
\end{figure}

In this paper, we focus on a typical 5G MIMO system operating in the time division duplexing (TDD) mode with OFDM modulation~\cite{larsson2014massive}. The channel estimation process consists of three main steps: (1) The user equipment (UE) transmits a predefined SRS pattern as specified by the 5G NR standard~\cite{3gpp.38.211} across \(N_{\text{UE}}\) antennas and over \(N_{K}\) out of \(N_{C}\) subcarriers.
(2) At the base station (BS), the signals received from \(N_{R}\) of the total \(N_{BS}\) antennas are analyzed to compute the LS estimate, which provides partial channel information in both domains. The neural network subsequently processes this input to reconstruct the complete channel matrix.
(3) Leveraging the estimated channel state information, the BS adjusts key communication parameters, such as modulation and coding schemes, to optimize communication performance.
This process is illustrated in Fig.~\ref{fig:background_channel_estimation}.

To mathematically describe the channel estimation process, we denote \(\mathbf{s}_i \in \mathbb{C}^{N_{\text{UE}} \times 1}\) as SRS sequence for \(i\)-th subcarrier. It is transmitted through the channel matrix \(\mathbf{H}_i \in \mathbb{C}^{N_{R} \times N_{\text{UE}}}\). The received signal \(\mathbf{y}_i \in \mathbb{C}^{N_{R} \times 1}\) at the BS is modeled as:
\vspace{-3pt}
\begin{equation}
    \mathbf{y}_i = \mathbf{A}_i \mathbf{H}_i \mathbf{s}_i + \mathbf{A}_i \mathbf{n}_i,
\vspace{-2pt}
\end{equation}
where \(\mathbf{A}_i \in \mathbb{C}^{N_{R} \times N_{R}}\) is the BS analog combiner matrix, and \(\mathbf{n}_i\) represents additive white Gaussian noise (AWGN). The effective channel matrix \(\mathbf{H}_{ei} \in \mathbb{C}^{N_{R} \times N_{{UE}}}\) is derived by combining \(\mathbf{A}_i \mathbf{H}_i\).

The statistical LS method minimizes the squared error between observed and predicted signals, yielding the estimation \(\mathbf{H}_{i}^{LS} = \operatorname*{arg\,min}_{\mathbf{H}} \|\mathbf{y}_i - \mathbf{H}_{ei} \mathbf{s}_i\|^2\). This expression can be simplified by the fact that the pilot sequences transmitted from each of the UE's antennas are designed to be orthogonal.

The orthogonality means the pilot signals do not interfere with each other, allowing the BS to separate the signal received from each transmit antenna. Mathematically, it allows the pilot to be modeled with a diagonal matrix \(\mathbf{s}_i \in \mathbb{C}^{N_{UE} \times N_{UE}}\). The well-known LS solution to the minimization problem is \(\mathbf{H}_{i}^{\text{LS}} = \mathbf{y}_i \mathbf{s}_i^{-1}\), which is equivalent to a direct element-wise division:
\begin{equation}
\label{form:ls}
\mathbf{H}{i}^{\text{LS}} = \mathbf{y}_i / \mathbf{s}_i.
\end{equation}

By concatenating the estimates across \(N_{K}\) subcarriers, the partial channel matrix \(\mathbf{H}^{LS} \in \mathbb{C}^{N_{K} \times (N_{R} \times N_{UE})}\) can be derived:
\vspace{-3pt}
\begin{equation}
    \mathbf{H}^{LS} = \left[\mathbf{H}_{1}^{LS}, \mathbf{H}_{2}^{LS}, \dots, \mathbf{H}_{N_{K}}^{LS}\right].
\vspace{-2pt}
\end{equation}

The LS estimation serves as an initial approximation of the channel state information and is subsequently fed into the neural network, which reconstructs the full channel matrix \(\mathbf{H}^{full} \in \mathbb{C}^{N_{C} \times (N_{BS} \times N_{UE})}\).

\section{DL-Based Channel Estimation}
\label{sec:network}

\begin{figure*}[ht!]
  \centering
  \includegraphics[width=\textwidth]{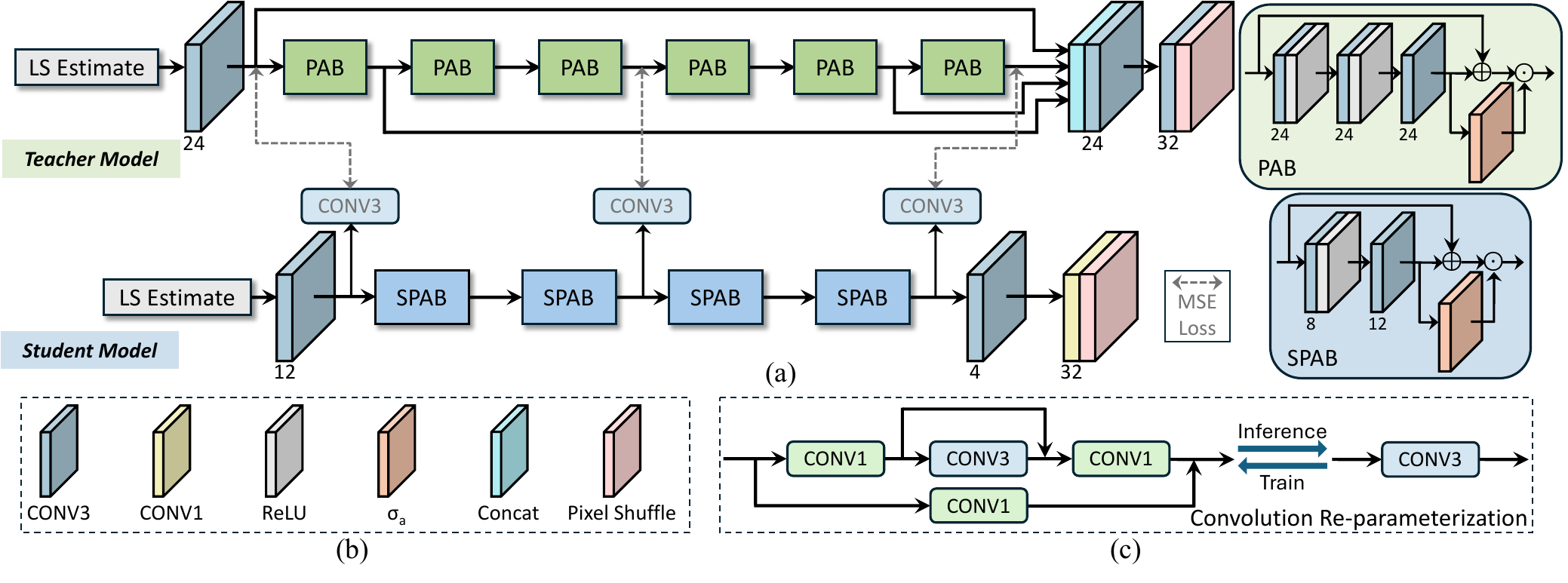}
  \vspace{-18pt}
  \caption{The architectures of the teacher and student model, along with the knowledge distillation process. (a) The internal structures of PABs and SPABs are shown in green and blue blocks, respectively, and the numbers below the layers indicate the number of output features. (b) A legend for symbols representing each layer type. (c) The illustration of the re-parameterization.}
  \vspace{-5pt}
  \label{fig:network_architecture}
\end{figure*}

In this section, we present the neural network architecture of the teacher model before knowledge distillation for frequency-spatial domain channel estimation, as shown in the upper part of Fig.~\ref{fig:network_architecture}(a).
The model extracts features through a cascaded convolution-based architecture and incorporates a parameter-free attention mechanism to enhance the ability to capture critical features and refine the channel matrix reconstruction.

\subsection{Convolution-based Feature Extraction}
The input LS estimations, represented as a dual-channel tensor of shape \(N_{K} \times (N_{R} \times N_{UE}) \times 2\), are processed using a sequential convolution-based architecture. The model begins with an initial \(3 \times 3\) convolutional layer, which transforms the real and imaginary input channels into a feature space with 24 channels, capturing essential low-level features.
At the core of the network is a sequence of six Parameter-Free Attention Blocks (PABs). Each PAB consists of three convolutional layers, two ReLU activation layers, and an attention layer. Together, these components allow the network to treat pixels unequally, focusing more on regions that carry significant information during gradient calculation and progressively extracting multi-level features.
The outputs of the final PAB, along with selected intermediate outputs from other layers, are concatenated to expand the teacher model's feature map to \(N_{K} \times (N_{R} \times N_{UE}) \times 96\), and then passed through an additional convolutional layer, which reduces the dimensionality back to 24 channels. This can effectively mitigate excessive information loss during the feature generation.
For the final step, the derived feature map is upsampled. This is achieved through a convolutional layer that transforms the feature map to a tensor of shape \(N_{K} \times (N_{R} \times N_{UE}) \times 32\) that serves as the input to the final pixel shuffle block. The pixel shuffle then rearranges the pixels into the full-resolution output of shape \(N_{C} \times (N_{BS} \times N_{UE}) \times 2\).
The number of output channels for each convolutional block is explicitly annotated in Fig.~\ref{fig:network_architecture}(a). The design choices of the architecture are empirically tuned to find the optimal balance between model capacity and our strict computational budget.

\subsection{Parameter-free Attention Mechanism}
Attention mechanisms have demonstrated significant potential in enhancing the wireless channel estimation performance by prioritizing critical signal features~\cite{yang2021deep, luan2023channelformer, zhou2023pay, zhou2025low}.
In estimation tasks, abrupt signal strength variations, interference patterns, and irregularities require focused processing to mitigate noise and blurring effects.
While traditional self-attention mechanisms offer such capabilities, their quadratic computational scaling with sequence length renders them impractical for large-scale MIMO systems~\cite{vaswani2017attention}.
To address this, we adopt a parameter-free attention mechanism inspired by advances in image super-resolution models~\cite{choe2019attention, yang2021simam, du2022parameter, shi2023parameter, wan2024swift}, which maintains effectiveness while avoiding computational overhead.

The proposed mechanism employs an origin-symmetric activation function $\sigma_a(x) = \text{sigmoid}(x) - 0.5$.
Within each PAB, the output of the final convolutional layer is passed through $\sigma_a$ to generate attention weights, which are applied to the sum of the feature map and the residual connection, yielding the block output:
\begin{equation}
    O_i = \sigma_a(H_i) \odot (H_i \oplus O_{i-1}),
\end{equation}
where \(H_i\) denotes the feature map, \(O_{i-1}\) is the block input of the \(i\)-th PAB, and \(\odot\) represents element-wise multiplication.
This design choice reflects our focus on achieving effective feature refinement through a lightweight mechanism, avoiding the high computational cost of more complex, parameter-heavy attention modules.

\subsection{Theoretical Analysis of Mechanism}
The effectiveness of the parameter-free attention mechanism can be theoretically validated through gradient backpropagation analysis. For simplicity, the residual connection is omitted first. For the \(i\)-th PAB processing features \(O_{i-1}\), the gradient updating weights without implementing attention is:
\begin{equation}
\frac{\partial \mathcal{L}}{\partial W_i} = \Pi \frac{\partial F^{(i)}_{W_i}(O_{i-1})}{\partial W_i},
\end{equation}
where \(F^{(i)}_{W_i}\) denotes the convolutional operations within the PAB, \(\mathcal{L}\) is the reconstruction loss \(\mathbb{E}[|\mathbf{H}^{full} - \hat{\mathbf{H}}^{full}|_2^2]\), and \(\Pi\) represents the gradient product from subsequent layers.
With attention, the gradient becomes:
\begin{equation}
\begin{split}
\frac{\partial \mathcal{L}}{\partial W_i}
&= \Pi \frac{\partial}{\partial W_i}\left[F^{(i)}_{W_i}(O_{i-1}) \odot \sigma_a(F^{(i)}_{W_i}(O_{i-1}))\right] \\
&= \Pi \frac{\partial F^{(i)}_{W_i}}{\partial W_i} \odot \left[H_i \odot \sigma'_a(H_i) + \sigma_a(H_i)\right],
\end{split}
\end{equation}
where the derivative of the activation function is given by:
\begin{equation}\sigma'_a(x) = \text{sigmoid}(x)(1 - \text{sigmoid}(x)) > 0 \ \forall x.\end{equation}

This derivative achieves a maximum value strictly less than \(0.25\) at \(H_i = 0\), and decreases monotonically as \(|H_i|\) increases. It introduces adaptive scaling to the weight updates based on the magnitude of the input features. For small feature values, \(\sigma'_a(H_i)\) remains relatively large, allowing for nuanced and sensitive updates in transitional or low-energy regimes.
In contrast, when \(|H_i|\) is large, indicating strong or high-energy features, the activation function \(\sigma_a(H_i)\) approaches \(\pm 0.5\) asymptotically, while the product \(H_i \odot \sigma'_a(H_i)\) tends toward zero.
This behavior ensures that channels with dominant features contribute a stable and bounded influence to the gradient, preventing uncontrolled amplification in high-energy regions and suppressing noise-induced perturbations.

Crucially, this mechanism selectively emphasizes regions with richer information content, where signal strength is moderate. These mid-range magnitudes correspond to areas where the derivative \(\sigma'_a(H_i)\) is non-negligible and the features are neither too noisy nor saturated. As a result, the gradient naturally focuses more on these informative zones, where meaningful distinctions in the data occur.

When residual connection is considered, the gradient expression incorporating attention modulation becomes:
\begin{equation}
\begin{split}
\frac{\partial \mathcal{L}}{\partial W_i}
&= \Pi \frac{\partial}{\partial W_i}\left[(F^{(i)}_{W_i}(O_{i-1}) + O_{i-1}) \odot \sigma_a(F^{(i)}_{W_i}(O_{i-1}))\right] \\
&= \Pi \frac{\partial F^{(i)}_{W_i}(O_{i-1})}{\partial W_i} \odot \left[(H_i + O_{i-1}) \odot \sigma'_a(H_i) + \sigma_a(H_i)\right],
\end{split}
\end{equation}
where the term \((H_i + O_{i-1})\) enables the model to retain historical information while adaptively refining estimates without impacting the attention mechanisms.

\section{Hardware-aware Model Compression}
\label{sec:optimization}
While the model presented in Section~\ref{sec:network} delivers excellent performance, its computation overhead and memory footprint may pose challenges for resource-constrained deployments. As a result, in this section, we propose a hardware-aware model compression framework aimed at reducing the model's complexity with marginal estimation performance degradation. The proposed pipeline consists of three crucial components: feature-based knowledge distillation, convolution re-parameterization, and quantization-aware training.

\subsection{Feature-based Knowledge Distillation}
The model described in Section~\ref{sec:network} is referred to as the teacher model, serving as the reference for feature-based knowledge distillation, which is an effective technique for transferring the capabilities of a large, high-capacity teacher model to a smaller, lightweight student model~\cite{hinton2015distilling}. To optimize the student model for hardware platforms, we replace the original PABs in the teacher model with four swift parameter-free attention blocks (SPABs).
As illustrated in Fig.~\ref{fig:network_architecture}(a), the SPAB is a significantly more lightweight version of the PAB; specifically, while the PAB consists of three convolutional layers operating on a 24-channel feature space, the SPAB is both shallower, containing only two layers, and narrower, operating on a 12-channel space, which is key to significantly lowering the model's complexity.
A major modification in the student model is the removal of feature concatenation. In the teacher model, concatenation introduces a latency bottleneck, requiring a convolutional layer to compress 96 features back into 24 channels.
By eliminating this step, we make a hardware-friendly design choice that avoids the use of excessive data buffers, thus reducing both on-chip and off-chip memory demands in on-board implementations.
The performance drop from this adjustment is mitigated through the distillation process, which has been shown to be effective in recovering model performance after the removal of residual connections~\cite{li2020residual, weng2024tailor}.

To ensure that the student model preserves the hierarchical feature extraction process of the teacher model and minimizes the information loss caused by the removal of feature concatenation, we define a combined distillation loss function that incorporates three components: hard loss, soft loss, and feature loss.
The hard loss measures the difference between the student model's final predictions and the ground truth, encouraging accurate output generation.
The soft loss compares the teacher's and student's outputs, allowing the student to benefit from the teacher's finer-grained estimations.
The feature loss aligns the intermediate feature representations of the teacher and student models to ensure the student mimics the teacher's hierarchical feature extraction process.
This approach has been widely utilized in many existing works~\cite{romero2014fitnets, zagoruyko2016paying, tung2019similarity}. A mean squared error (MSE) alignment loss is applied to the middle-layer outputs of both models. However, because the dimensions of the teacher and student features differ, \(3 \times 3\) convolutional layers are used to project the student's intermediate outputs into a compatible feature space. Overall, all three losses are combined based on the following equation to achieve a balanced performance:
\vspace{-3pt}
\begin{equation}
    \mathcal{L}_{\text{distillation}} = \alpha \times \mathcal{L}_{\text{hard}} + \beta \times \mathcal{L}_{\text{soft}} + \gamma \times \mathcal{L}_{\text{feature}}.
\vspace{-2pt}
\end{equation}

\subsection{Convolution Re-parameterization}
As the student model must retain sufficient capacity to effectively mimic the feature extractor of the teacher model, its total number of parameters cannot be excessively reduced.
To enhance the estimation performance of the student model during knowledge distillation, we employ convolution re-parameterization techniques before the training.
As illustrated in Fig.~\ref{fig:network_architecture}(c), during training, we use a composite block consisting of one \(3 \times 3\) convolutional layer and three \(1 \times 1\) convolutional layers. This configuration can be subsequently replaced by an equivalent, simplified \(3 \times 3\) convolutional layer during inference.
This transformation significantly reduces the model size and parameter count, while maintaining representational power, which is a strategy that has demonstrated strong potential for improving algorithmic efficiency~\cite{wan2024swift, ding2021repvgg}.

This technique offers two key benefits. First, it introduces additional complexity to the model structure during training, thus allowing it to more closely mimic the teacher model’s behavior to enhance the performance.
Second, the re-parameterization facilitates the inclusion of consecutive skip connections, which mitigate potential information loss caused by the removal of feature concatenation. Overall, this technique strikes a balance between training expressiveness and inference efficiency, making it well-suited for knowledge distillation scenarios where compactness and performance must coexist.

\begin{figure*}[t]
  \centering
  \includegraphics[width=\linewidth]{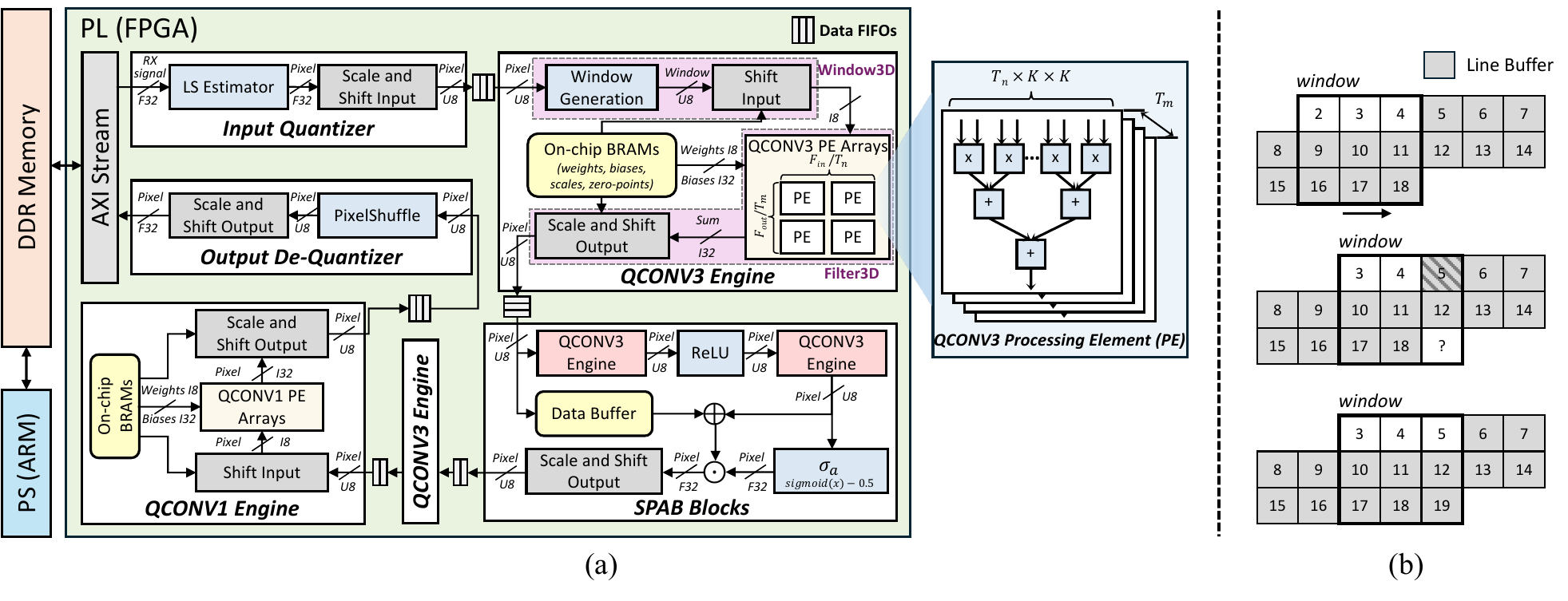}
  \vspace{-18pt}
  \caption{(a) The hardware accelerator architecture targeting the FPGA platforms. (b) An illustration depicting the window generation process within QCONV3 engines.}
  \label{fig:hardware_architecture}
  \vspace{-5pt}
\end{figure*}

\subsection{Quantization-aware Training}
Quantization is a key technique for reducing the computational cost and memory requirements of neural networks by converting floating-point representations of the weights and biases to lower-bit integer formats~\cite{guo2017angel, chang2021mix, sun2022film}.
Quantization-aware training (QAT) improves deployment performance by simulating integer representations during training, allowing the model to adapt to low-precision constraints, unlike post-training quantization (PTQ), which applies calibration only after training.

After training the student model with knowledge distillation, the model is further trained while all the weights are quantized to 8-bit integers, and biases are represented as 32-bit integers.
Activation functions such as ReLU and Sigmoid are excluded from QAT, which means that quantization nodes are not applied to the non-linear operators themselves during the training phase.
Apart from weights, input data is similarly quantized with the following equation:
\vspace{-3pt}
\begin{equation}
    x_q = \text{round}\left(\frac{x}{\text{scale}_x}\right) + \text{zero\_point}_x,
\vspace{-2pt}
\end{equation}
where \(x\) is the original floating-point input, \(x_q\) is the quantized unsigned 8-bit integer, \(\text{scale}_x\) is the scaling factor, and \(\text{zero\_point}_x\) is the offset ensuring the representation of both positive and negative values.

In convolutional layers, the integer-only multiplications are performed on integer-based inputs and weights. The multiplication results are stored using 32 bits. In the end, they should be re-quantized back to 8-bit with the following equation:
\vspace{-3pt}
\begin{equation}
\label{form:zq}
    z_q = \text{round}\left(\frac{\text{scale}_x \times x_s \times \text{scale}_w \times w_s}{\text{scale}_z}\right) + \text{zero\_point}_z,
\vspace{-2pt}
\end{equation}
where \(x_s = x_q - \text{zero\_point}_x\) and \(w_s = w_q - \text{zero\_point}_w\) are integer representations adjusted by their respective zero-points.
Ultimately, QAT minimizes the performance drop while fully leveraging the computational advantages of quantization.

\section{Hardware Accelerator Implementation}
\label{sec:hardware}
In this section, we detail a novel hardware accelerator architecture implementing the DL-based channel estimation algorithm derived from the model compression with a fine-grained pipeline architecture, which minimizes the overall latency within resource utilization limitations.

\subsection{Design Methodology and Notations}
The hardware design is delivered with HLS, which converts the abstract behavioral specification into a register-transfer level (RTL) structure, ensuring an optimized hardware architecture while maintaining flexibility.
The proposed architecture is inherently platform-independent due to its reliance on modular design principles. By leveraging HLS, the system avoids tight coupling to vendor-specific IP cores, enabling straightforward deployment on various FPGA platforms.

Fig.~\ref{fig:hardware_architecture}(a) outlines the architecture of the proposed hardware accelerator. All the key components are mapped onto the programmable logic (PL), facilitating parallel processing and maximizing throughput. Meanwhile, control and scheduling are managed by the processing system (PS), an ARM processor.
The hardware IP is designed with AXI protocol, ensuring compatibility with a wide range of SoCs and facilitating easy integration into various systems. Direct memory access (DMA) is employed for high-throughput data transfers between the PS, PL, and external DDR memory.

To clearly describe the numerical formats used in the design, which include fixed-point and integer numbers with various bit widths, we adopt standardized abbreviations throughout this section. For example, INT8/I8 represents signed 8-bit integers, UINT8/U8 denotes unsigned 8-bit integers, and FIX32/F32 refers to 32-bit fixed-point numbers (with 7 bits allocated for the integer part). Floating-point numbers are not used in hardware design to conserve hardware resources, as fixed-point numbers provide sufficient accuracy. Additionally, Table~\ref{tab:symbol} provides a comprehensive list of other symbol notations and their explanations.

\begin{table}[t]
\centering
    \begin{minipage}{\linewidth}
    \centering
    \caption{Symbol Notations in Section~\ref{sec:hardware}}
    \label{tab:symbol}
    \begin{tabular}{c|l}
    \toprule
        \textbf{Symbol}& \textbf{Description}\\ \midrule
        $H$            & Height of the LS-estimate (frequency domain) \\
        $W$            & Width of the LS-estimate (spatial domain) \\
        $C_{in}$       & Number of channels of the LS-estimate (\ie 2) \\
        $C_{out}$      & Number of channels of the final output (\ie 2) \\
        $F_{in}$       & Number of input channels of convolutional layer \\
        $F_{out}$      & Number of output channels of convolutional layer \\
        $K$            & Kernel size of the convolution \\
        $r$            & Scale factors of both domains (\ie 4) \\
    \bottomrule
    \end{tabular}
\end{minipage}
\vspace{-5pt}
\end{table}

\subsection{Fine-Grained Pipeline Architecture}
Given the 5G frame structure, where SRS transmissions occur every \SI{1}{ms}, the processing delay for each SRS must remain below this threshold to meet real-time requirements.
To reduce overall latency, the design utilizes a fine-grained pipeline architecture that breaks down the algorithm into small, independent processing blocks connected through HLS streams, which are essentially data FIFOs in the generated RTL design.
The use of streaming components minimizes intermediate buffering, reducing on-chip memory overhead while maintaining deterministic dataflow.
The pipeline allows efficient processing by breaking computations into minimal stages that run concurrently, overlapping data transfers and computations, although this may require more hardware resources.
Each block processes the data as soon as it becomes available, thereby enabling a continuous dataflow across the pipeline. Thus, the overall latency is dictated by the slowest block rather than the cumulative latency of all blocks. This architecture introduces both intra-block parallelism and inter-block parallelism.
Furthermore, the modularity of the fine-grained pipeline ensures scalability, so that individual blocks can be modified or replaced without redesigning the entire system, making the architecture adaptable to evolving algorithmic requirements and increasing complexity.

\subsection{LS Estimator Block}
The computation of LS estimation, as defined in Equation~\ref{form:ls}, is relatively straightforward and computationally efficient, primarily involving complex number division between the received data samples and the SRS patterns.
To minimize communication delays in accessing memory, the transmitted SRS is pre-stored in the on-chip block RAMs (BRAMs), which act as a local cache.
To ensure efficient hardware implementation of complex division, both the real and imaginary parts of the patterns are normalized by the squared magnitude of the SRS \(s_{real}^2 + s_{imag}^2\) before being cached. This approach avoids direct division operations by using pre-computed values for the denominator.
Both the input and output data of the block are represented as FIX32 numbers. The output data is then scaled and shifted to UINT8 before being passed to the subsequent QCONV3 engine.

\subsection{QCONV3 Engine Blocks}
The most crucial block in the accelerator design is the quantized convolutional layer with \(3 \times 3\) kernel, which occupies most resources and acts as the latency bottleneck. Thus, the design of an efficient QCONV3 engine to meet both the latency and resource requirements presents a critical challenge. The engine processes input data as a UINT8 stream in depth-first order and outputs a UINT8 stream as quantized feature maps. The designed engine consists of two modules, Window3D and Filter3D, as depicted in Fig.~\ref{fig:hardware_architecture}(a). The Window3D is responsible for window generation and input shifting. This module creates a local buffer of pixels, enabling data reuse. The Filter3D houses an array of Processing Elements (PEs) that handle convolution operations.

\subsubsection{Window3D Module}
The Window3D in the convolution engine efficiently manages a sliding \(3 \times 3\) window over input data, preparing it for the subsequent filtering operation. It leverages two local buffers cached in BRAM. The first one is the line buffer, a rolling buffer that stores \(K-1\) rows of input pixels.
Another one is the window buffer that stores \(3 \times 3 \times F_{in}\) values, representing the active sliding window for each input channel. Fig.~\ref{fig:hardware_architecture}(b) provides a visual depiction of the loop that updates both buffers.
The window buffer shifts horizontally to accommodate the latest pixel, simulating a sliding window, while the line buffer updates afterward with the help of column and channel pointers to track the progress.

After the initial ramp-up phase, where enough data has populated the line buffer and window buffer, the function begins populating the output window.
Two data pre-processing steps are performed here. The data padding ensures zero padding for window regions that exceed the boundaries. The input shifting follows \(x_{s} = x_q - \text{zero\_point}_x\) to prepare data for subsequent quantized multiplications. The pre-processed INT8 window data is then written to the output stream, ready for retrieval by later layers.

\subsubsection{Filter3D Module}
\begin{algorithm}[t]
\caption{Accelerated Convolution Operation}
\label{alg:tiled_conv}
\begin{algorithmic}[1]
\footnotesize

\Statex \textbf{Input:} $T_n, T_m$: tile sizes, $window\_stream$: input stream
\Statex \textbf{Output:} $pixel\_stream$: output stream

\State \textbf{Initialize} $sum[F_{out}]$ to hold multiplication results
\State \textbf{Initialize} $partial\_sum[T_m]$ to hold partial sums
\State \textbf{for} {$y = 0$ to $H-1$} \textbf{do}
\State \textbf{for} {$x = 0$ to $W-1$} \textbf{do}
\State \hspace{\algorithmicindent} {$window[\ ] \gets window\_stream$} \Comment{Read the input window}
\State \hspace{\algorithmicindent} {$sum[\ ] \gets biases[\ ]$} \Comment{Initialize sums with bias}
\State \hspace{\algorithmicindent} \textbf{for} {$to = 0$ to $F_{out}-1$ by $T_m$ \textbf{do}
\State \hspace{\algorithmicindent} \textbf{for} {$ti = 0$ to $F_{in}-1$ by $T_n$} \textbf{do}
\State {\#pragma HLS PIPELINE}
\State \hspace{\algorithmicindent} \hspace{\algorithmicindent} $partial\_sum[\ ] \gets 0$ \Comment{Reset partial sums to 0s}
\State \hspace{\algorithmicindent} \hspace{\algorithmicindent} \textbf{for} {$rr = 0$ to $K-1$} \textbf{do}
\State \hspace{\algorithmicindent} \hspace{\algorithmicindent} \textbf{for} {$cc = 0$ to $K-1$} \textbf{do}
\State \hspace{\algorithmicindent} \hspace{\algorithmicindent} \textbf{for} {$too = 0$ to $T_m-1$} \textbf{do}
\State \hspace{\algorithmicindent} \hspace{\algorithmicindent} \textbf{for} {$tii = 0$ to $T_n-1$} \textbf{do}
\State {\#pragma HLS UNROLL}
\State \hspace{\algorithmicindent} \hspace{\algorithmicindent} \hspace{\algorithmicindent} $partial\_sum[too]$ += $window[ti+tii][rr][cc]$ 
\State \hfill $ \times weights[ti+tii][rr][cc][to+too]$

\State \hspace{\algorithmicindent} \hspace{\algorithmicindent} $sum[\ ]$ += $ partial\_sum[\ ]$ \Comment{Accumulate partial sums}

\State \hspace{\algorithmicindent} \textbf{for} {$fo = 0$ to $F_{out}-1$} \textbf{do}
\State \hspace{\algorithmicindent} \hspace{\algorithmicindent} $res \gets \text{scale\_shift}\left(sum[fo], scales, zero\_points\right)$
\State \hspace{\algorithmicindent} \hspace{\algorithmicindent} {$res \to pixel\_stream$} \Comment{Write re-quantized output}
}
\end{algorithmic}
\end{algorithm}
Filter3D focuses on the filtering operations on the sliding window. To facilitate its performance, four loop optimization techniques—loop unrolling, loop tiling, loop interchange, and loop pipelining—have been considered to optimize the computation patterns and efficient mapping onto hardware. The optimized convolution operation is detailed in Algorithm~\ref{alg:tiled_conv}.

Loop tiling plays a critical role in determining both parallel processing efficiency and memory utilization.
Previous works~\cite{liang2019evaluating, bai2020unified, ujjainkar2023imagen, ma2017optimizing, zhang2015optimizing} have extensively explored loop tiling strategies, showing that different tiling strategies offer varying opportunities for data reuse.
In our design, we apply loop tiling to both the output channels (line 7 in Algorithm~\ref{alg:tiled_conv}) and input channels (line 8 in Algorithm~\ref{alg:tiled_conv}).
Tiling the output channels into \(T_m\) tiles partitions the workload into smaller segments and reduces the number of partial sums that need to be computed in concurrency. Similarly, tiling the input channels into \(T_n\) tiles ensures that only a subset of the input feature map is loaded and processed at a time. This dual-tiling strategy leads to an optimal balance between performance and resource efficiency. The values of \(T_m\) and \(T_n\) determine the total number of PEs required, as described by the equation:
\vspace{-5pt}
\begin{equation}
    \text{Number of PEs} = \frac{F_{out}}{T_m} \times \frac{F_{in}}{T_n}.
\vspace{-3pt}
\end{equation}

For each PE, one partial sum is computed, while all the operations inside the PE are fully unrolled by inserting an HLS pragma (line 15 in Algorithm~\ref{alg:tiled_conv}). The PE predominantly uses DSPs to perform the multiply-accumulate operation, where each shifted pixel is multiplied by the corresponding INT8 weight, and the result is accumulated into the INT32 partial sum. The internal structure of PE is shown in Fig.~\ref{fig:hardware_architecture}(a). These partial sums are then accumulated into the sum array initialized by the biases for each output channel.

Loop interchange is applied to change the order of computation, reordering the loops to a more efficient order: first iterating over all kernel dimensions, followed by output channels, and then input channels (lines 11-14 in Algorithm~\ref{alg:tiled_conv}). This reordering improves the efficiency of data access for the weights and input windows, as the weights matrix can now be partitioned entirely across kernel dimensions. It reduces memory access latency and improves spatial locality, as consecutive iterations now operate on nearby elements in memory.

Finally, the loops are further optimized by applying the pipeline directive to the outer loop (line 9 in Algorithm~\ref{alg:tiled_conv}).
This strategy targets an optimal Initiation Interval (II) of 1 to maximize throughput.
Achieving an II of 1 inherently necessitates the full unrolling of the inner loops (line 15 in Algorithm~\ref{alg:tiled_conv}) by the HLS tool, creating sufficient parallel hardware instances to process new data every clock cycle without stalling.
This design choice accepts higher resource utilization as a necessary trade-off to strictly satisfy the sub-millisecond latency requirement.
In the end, the result pixel value undergoes scaling and shifting as described in Equation~\ref{form:zq} before writing to the output stream.

\subsection{SPAB Blocks}
The implementation of the SPAB blocks incorporates two QCONV3 engines connected in sequence, with a ReLU activation stream situated between them. For the ReLU activation, each UINT8 pixel value is compared to the zero-point of the first convolution layer's output. If a pixel value exceeds the zero-point, it remains unchanged; otherwise, it is set to the value of the zero-point, mimicking the behavior of the ReLU function in floating-point systems.

The skip connection in the SPAB block is implemented by duplicating the input stream with a bypass data FIFO, which is realized as a BRAM. During computation, one value is extracted from the bypass FIFO, while another is taken from the output stream of the second convolution layer.
The latter undergoes a custom activation function \(\sigma_a\) implemented as a precomputed lookup table stored in BRAM.
To perform the lookup, the incoming UINT8 pixel value is first de-quantized into its corresponding high-precision FIX32 representation. This FIX32 value is then used to index the lookup table, which contains \(512\) pre-computed points for inputs in the range \((-3, 3)\), with each entry representing the activation result in FIX32 format.
To preserve mathematical fidelity during the attention calculation, both the bypassed input and the second convolution output are converted to FIX32 format, summed, and multiplied by the calculated attention weights. Each result pixel value is re-quantized to the UINT8 format in the end.

\subsection{PixelShuffle Block}
Despite requiring no explicit computation, the PixelShuffle block presents significant challenges for hardware implementation due to the high latency incurred by the extensive data buffering required in a naive, non-pipelined implementation, leading to suboptimal hardware performance. The input of the block is the feature map generated by the upsampling layer \(O_{up} \in \mathbb{R}^{H \times W \times \left(C_{out} \times r^2\right)}\). During the pixel shuffle process, the feature dimension of each pixel is reshaped and transposed into a spatial region of size \(r \times r\) across \(C_{out}\) channels, resulting in a final feature map \(O_{final} \in \mathbb{R}^{(r \times H) \times (r \times W) \times C_{out}}\).

As each pixel only contributes to a small spatial region within the output, a complete row of the final feature map is formed by concatenating \(W\) such regions in sequence.
This means the unaccelerated pixel shuffle algorithm requires a large, monolithic line buffer (typically implemented in BRAMs) of at least \(W \times \left(C_{out} \times r^2\right)\) pixels to maintain the correct channel-width-height output order, and two clock cycles are needed for each pixel's reading and writing operations. It has already been a significant bottleneck in our design, even surpassing the latency of the QCONV3 engine.

\begin{figure}[t]
  \centering
  \includegraphics[width=0.98\linewidth]{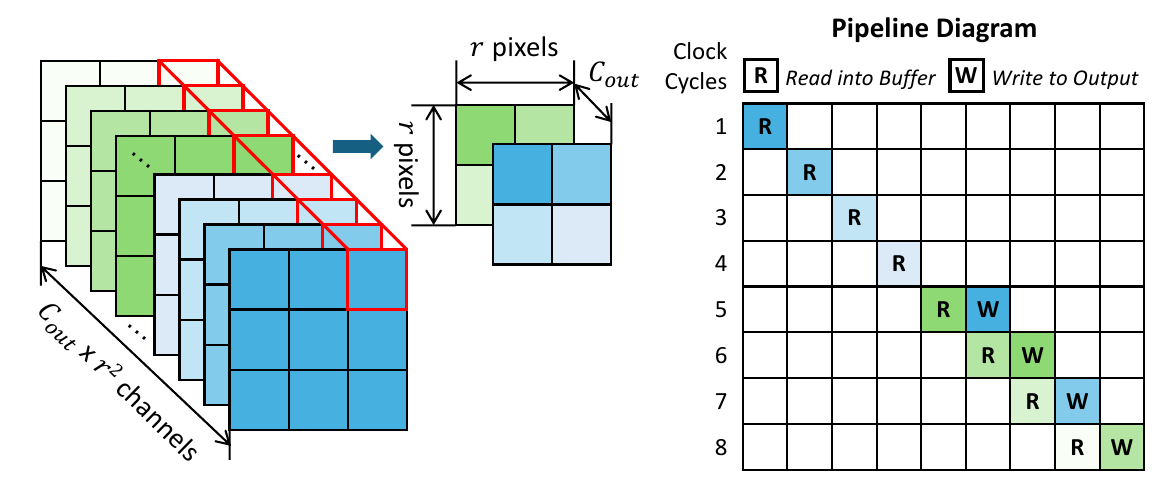}
  \vspace{-5pt}
  \caption{Pipeline diagram for pixel shuffle acceleration.}
  \label{fig:hardware_pixelshuffle}
  \vspace{-5pt}
\end{figure}

To mitigate this bottleneck, we propose a pipelined structure for the pixel shuffle operation. This new architecture does not eliminate intermediate buffering, but rather replaces the single, large BRAM-based buffer with a more efficient form of buffering: a system of smaller, distributed FIFOs and registers.
Fig.~\ref{fig:hardware_pixelshuffle} illustrates the pipeline diagram of the proposed acceleration mechanism. The reading and writing operations for the first row's pixels in each region are executed in parallel. Subsequent pixels from other scaled rows are temporarily buffered as they need to wait until one entire row of the final feature map is filled out. Once the buffering is complete, the function outputs the remaining pixels in the correct sequence. Simultaneously, the process begins reading the new pixels from the input feature map, allowing for deeper parallelism. This approach significantly reduces the latency introduced by the operation at the cost of higher complexity in logic control.

\section{Evaluation}
\label{sec:evaluation}

In this section, we evaluate the performance of \sysname through a co-design perspective, evaluating both model accuracy and efficiency, alongside the latency and resource utilization of the dedicated hardware accelerator. Through extensive experiments and comparative analyses, we aim to validate the efficacy of our approach in achieving superior performance while meeting stringent efficiency demands.

\subsection{Experimental Settings}
\subsubsection{Dataset Synthesis}
\begin{table}[t!]
\centering
\caption{5G MIMO System and Dataset Parameter Settings.}
\label{tab:sim_params}
\renewcommand{\arraystretch}{1.1}
\begin{tabular}{l|c}
\toprule
\textbf{Parameter} & \textbf{Value} \\
\midrule
\multicolumn{2}{c}{\textit{5G MIMO System Parameters}} \\
\midrule
Carrier frequency \(f_c\) & 28 GHz \\
Numerology \(\mu\) & 3 \\
Subcarrier spacing & 120 kHz \\
SRS period \(T\) & 1 ms \\
Number of BS antennas \(N_{BS}\) & 64 \\
Number of UE antennas \(N_{UE}\) & 2 \\
Number of resource blocks \(N_{RB}\) & 36 \\
Total subcarriers \(N_{C}\) & 432 \\
Spatial scale factor \(N_{BS}/N_{R}\) & 4 \\
Frequency scale factor \(N_{C}/N_{K}\) & 4 \\
\midrule
\multicolumn{2}{c}{\textit{Dataset Generation Parameters}} \\
\midrule
Training Channel model & CDL-B \\
Testing Channel model & CDL-A/B/C/D/E \\
Training UE velocity (km/h) & 5, 15, 30, 45, 60 \\
Testing UE velocity (km/h) & 5, 20, 40, 60, 80, 100, 120 \\
Training SNR levels (dB) & 5, 10, 15, 20 \\
Testing SNR levels (dB) & 6, 10, 14, 18, 22, 26, 30 \\
Number of training samples & 12,800 \\
Number of validation samples & 3,200 \\
Number of testing samples & 15,680 \\
\bottomrule
\end{tabular}
\vspace{-10pt}
\end{table}
The channel estimation dataset was constructed using MATLAB’s 5G Toolbox, which provides highly accurate channel modeling and is widely adopted in similar research endeavors~\cite{soltani2019deep, li2019deep, luan2021low, luan2023channelformer, yang2021deep, sharma2024low, zhou2023pay, zhou2025low}. All system and dataset generation parameters are summarized in Table~\ref{tab:sim_params}.

The setup comprises a BS MIMO system with a uniform linear array at the base station and multiple antennas at the UE. The frequency domain parameters, including the number of resource blocks and subcarrier spacing, are configured according to the 5G NR standard for millimeter-wave scenarios using the clustered delay line (CDL) channel models. The SRS symbol is configured as the last symbol of one slot per sub-frame and is transmitted periodically. We define spatial and frequency scale factors as the ratio of active-to-total antennas and active-to-total subcarriers, respectively.
As a result, the output derived from the LS estimator is processed as a real-valued tensor \(\boldsymbol{H}^{\text{LS}} \in \mathbb{R}^{108 \times 32 \times 2}\), which the proposed DL-based method subsequently upscales to reconstruct the full-resolution channel matrix \(\boldsymbol{H}^{\text{full}} \in \mathbb{R}^{432 \times 128 \times 2}\).

To ensure robustness and evaluate the model's generalization capability, the dataset was generated under diverse channel conditions. The training dataset was generated using the channel model CDL-B, which represents a common yet challenging urban scenario, characterized by a rich, non-line-of-sight multipath environment that provides diverse features for the model to learn.
AWGN is added at the receiver for each SNR level, while UE velocity determines the maximum Doppler shift.
In contrast, the extensive test dataset was created not only with varying unseen combinations of SNR and velocities but also using a diverse range of channel models: the Non-Line-of-Sight (NLoS) profiles CDL-A, B, and C, and the Line-of-Sight (LoS) profiles CDL-D and E.
This methodology allows us to evaluate the model's performance on both seen and entirely unseen channel profiles, providing a robust measure of its real-world applicability.

\subsubsection{Model Training Protocol}
The teacher model was trained using the Adam optimizer with an initial learning rate of \(2 \times 10^{-4}\), decaying by a factor of \(0.9\) every \(40\) epochs to stabilize convergence. A batch size of \(16\) was selected to balance gradient variance reduction with memory constraints inherent to high-dimensional channel data. Training proceeded for a maximum of \(400\) epochs, with early stopping activated after \(20\) consecutive validation periods without performance improvement, mitigating overfitting risks.

The model compression framework begins with knowledge distillation, where the student model is trained using logits and intermediate feature maps from the teacher model. During this phase, the initial learning rate was set as \(2 \times 10^{-4}\) and training was constrained to a maximum of \(200\) epochs.
To equally balance contributions from three loss components \(\mathcal{L}_{\text{hard}}\), \(\mathcal{L}_{\text{soft}}\) and \(\mathcal{L}_{\text{feature}}\), we performed empirical scale normalization.
Specifically, average magnitudes of individual loss terms were computed over a subset of the training data to quantify their inherent imbalances. Weights \(\alpha\), \(\beta\), and \(\gamma\) were then assigned inverse proportionality to these baselines, after which they were chosen as \(\alpha = 1\), \(\beta = 10\), and \(\gamma = 2\).
Following distillation and convolution re-parameterization, per-channel quantization is integrated into the final stage of training. Here, weight updates are regularized to approximate low-precision representations. The QAT phase utilizes a reduced learning rate of \(5 \times 10^{-4}\) and spans up to \(100\) epochs, ensuring convergence while mitigating catastrophic forgetting of distilled knowledge.

\subsection{Evaluation on Algorithm Performance}

\begin{table}[t]
\centering
\caption{Complexity comparison with baseline approaches.}
\label{tab:evaluation_baseline}
\renewcommand{\arraystretch}{1.1}
\begin{tabular}{l|c|c|c|c}
\toprule
\textbf{Model} & \makecell{\textbf{NMSE} \\ \textbf{(-dB)}} & \makecell{\textbf{Parameters} \\ \textbf{(K)}} & \makecell{\textbf{FLOPs} \\ \textbf{(M)}} & \makecell{\textbf{Attention} \\ \textbf{Mechanism}} \\
\midrule
\textbf{\textit{SwiftChannel} (ours)} & \textbf{6.833} & \textbf{7.816} & \textbf{26.57} & $\checkmark$ \\ \midrule
ChannelNet~\cite{soltani2019deep} & 0.002 & 675.1 & 37,327 & $\times$ \\
ReEsNet~\cite{li2019deep} & 5.712 & 52.47 & 195.6 & $\times$ \\
I-ResNet~\cite{luan2021low} & 3.633 & 6.190 & 63.54 & $\times$ \\
Channelformer~\cite{luan2023channelformer} & 7.600 & 1,146,834 & 9,740 & $\checkmark$ \\
LSiDNN~\cite{sharma2024low} & 4.690 & 5,751 & 5.640 & $\times$ \\
FSRCNN~\cite{dong2016accelerating} & -0.083 & 12.81 & 43.08 & $\times$ \\
\bottomrule
\end{tabular}
\vspace{-10pt}
\end{table}

\begin{figure*}[t]
  \centering
  \includegraphics[width=\linewidth]{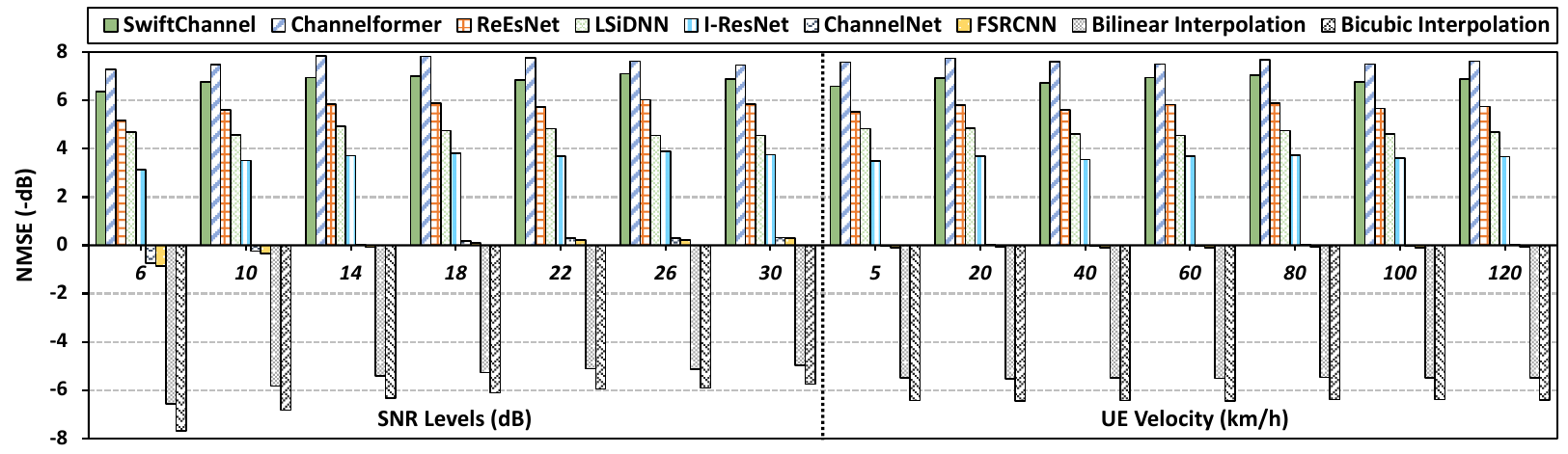}
  \vspace{-15pt}
  \caption{Model performance comparison with baselines across different settings.
  Performance is averaged over the full range of another variable for each data point (e.g., the result at each SNR is an average over all tested UE velocities, and vice versa).}
  \label{fig:evaluation_baseline}
  \vspace{-8pt}
\end{figure*}

To validate the performance of our proposed algorithm \sysname, we conduct extensive evaluations across unseen SNR levels and UE velocities.
For the initial set of evaluations, all algorithms are evaluated on the CDL-B channel profile to establish a baseline performance on the seen channel model. Subsequently, to rigorously assess the proposed method's generalization capabilities, we will conduct a final evaluation where we evaluate its performance on several entirely unseen channel models.
Channel estimation accuracy is evaluated via Normalized Mean Squared Error (NMSE) with the formula of \(\frac{\mathbb{E}\left[\|\mathbf{H}^{full} - \hat{\mathbf{H}}^{full}\|_F^2\right]}{\mathbb{E}\left[\|\mathbf{H}^{full}\|_F^2\right]},\)
where \(\mathbf{H}^{full}\) represents the ground-truth channel matrix and \(\hat{\mathbf{H}}^{full}\) denotes the estimated counterpart.  The computational complexity of the algorithm design is measured in terms of the number of learnable parameters and floating-point operations per second (FLOPs).

\subsubsection{Comparison with Baseline Approaches}
Comparative analyses involve six DL-based baselines including ChannelNet\cite{soltani2019deep}, ReEsNet \cite{li2019deep}, I-ResNet\cite{luan2021low}, Channelformer \cite{luan2023channelformer}, LSiDNN\cite{sharma2024low}, FSRCNN \cite{dong2016accelerating}, and traditional interpolators.

To ensure a fair and rigorous comparison, the architecture of each DL-based baseline was adapted to our system model.
We strictly followed the architectural definitions of the original models, adjusting structural hyperparameters, such as embedding dimensions for Transformer-based architectures or upsampling parameters for convolutional models, where necessary to align with the specific spatial-frequency resolution of our dataset.
Specifically, the non-pruned offline version of ChannelFormer~\cite{luan2023channelformer} was selected to evaluate the model's performance upper bound, while its complexity could be further reduced using the pruning methods introduced in~\cite{luan2023channelformer} with the performance trade-off.
Finally, all adapted baseline models were retrained from scratch using the similar protocol as our proposed model, with hyperparameters individually optimized for each architecture.

The resulting complexity comparison is presented in Table~\ref{tab:evaluation_baseline}, where our proposed model demonstrates high efficiency, with only \(7.816\,\mathrm{K}\) parameters and a computational cost of \(26.57\,\mathrm{M}\) FLOPs.
The performance of each algorithm across different settings on the seen channel model (CDL-B) is presented in Fig.~\ref{fig:evaluation_baseline}.
Our results demonstrate that \sysname achieves an NMSE of \(-6.833\,\mathrm{dB}\), outperforming most baselines except Channelformer~\cite{luan2023channelformer}, which has a slightly better NMSE of \(-7.600\,\mathrm{dB}\).
However, \sysname is significantly more efficient, requiring far fewer computational resources than Channelformer, which demands \(9.74\,\mathrm{G}\) FLOPs and over \(1.15\,\mathrm{G}\) parameters.
This highlights the excellent performance that the algorithm has achieved by employing a parameter-free attention mechanism without introducing a large computational overhead like the self-attention mechanism.
ReEsNet~\cite{li2019deep}, while competitive in NMSE, incurs \(195.6\,\mathrm{M}\) FLOPs and \(52.5\,\mathrm{K}\) parameters, rendering it less practical for latency-constrained scenarios.
Other methods significantly lag behind the proposed algorithm in accuracy and are less efficient.
Overall, \sysname offers an excellent trade-off between estimation performance and computational efficiency, positioning it as a strong candidate for channel estimation tasks.

\subsubsection{Ablation Study on Model Performance}
\begin{figure*}  
  \centering 
  \includegraphics[width=\linewidth]{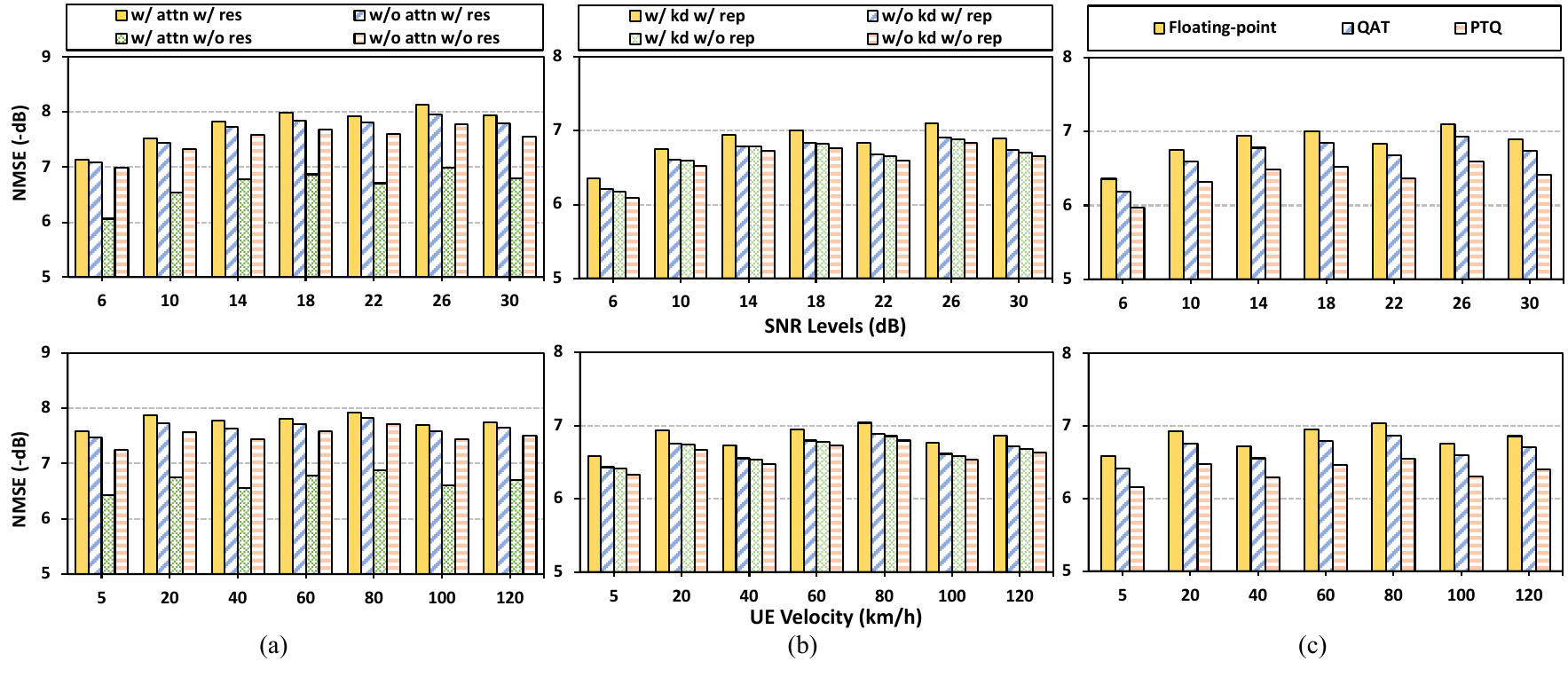}  
  \vspace{-22pt}  
  \caption{(a) Ablation study of parameter-free attention mechanism (attn) and residual connection (res). (b) Ablation study of knowledge distillation (kd) and re-parameterization (rep). (c) Ablation study of model quantization.}
  \label{fig:evaluation_ablation_study}  
  \vspace{-10pt}  
\end{figure*}

To evaluate the contributions of the parameter-free attention mechanism and residual connections in the model architecture, we conducted an ablation study by systematically ablating these components and assessing the teacher model’s performance under varied configurations prior to knowledge distillation.
As illustrated in Fig.~\ref{fig:evaluation_ablation_study}(a), removing the parameter-free attention mechanism leads to a noticeable degradation in NMSE, highlighting its critical role in adaptively refining spatial-frequency correlations.
Disabling residual connections exacerbates a much higher performance loss than removing attention, consistent with findings in~\cite{wan2024swift}, as the model struggles to capture deep hierarchical features effectively, and gradient propagation bottlenecks hinder deep hierarchical feature extraction.
Configurations lacking both components also yield bad performance, underscoring the synergistic importance of these design choices in achieving robust channel estimation.

In Fig.~\ref{fig:evaluation_ablation_study}(b), we evaluate the synergistic impact of knowledge distillation and re-parameterization by comparing the final compressed model against three alternative configurations: (1) direct training of the student model without distillation, (2) distillation applied to a student model without re-parameterization, and (3) direct training of a non-re-parameterized student model.
Training the student model independently yields suboptimal NMSE, demonstrating its inherent inability to learn complex channel features from raw data without teacher guidance.
Disabling re-parameterization is equivalent to reducing architectural complexity during training, which similarly degrades accuracy due to the loss of structural expressiveness.
These results validate that the integration of both techniques enables the student model to assimilate teacher knowledge while retaining sufficient architectural capacity during training, ultimately achieving a compact, deployment-ready architecture post-optimization.

Quantization remains a pivotal component of model compression. As shown in Fig.~\ref{fig:evaluation_ablation_study}(c), a comparative analysis of QAT and PTQ consistently favors QAT across all test settings. While quantization inherently incurs minor accuracy degradation, QAT mitigates this by recalibrating parameter distributions and optimizing zero-point and scale factor selections during low-precision adaptation.
This iterative refinement ensures hardware-compatible weight representations with minimal performance loss, rendering the trade-off between model size reduction and accuracy preservation practically viable.

\subsubsection{Effectiveness of Compression Techniques}

\begin{figure}[t]
  \centering
  \includegraphics[width=\linewidth]{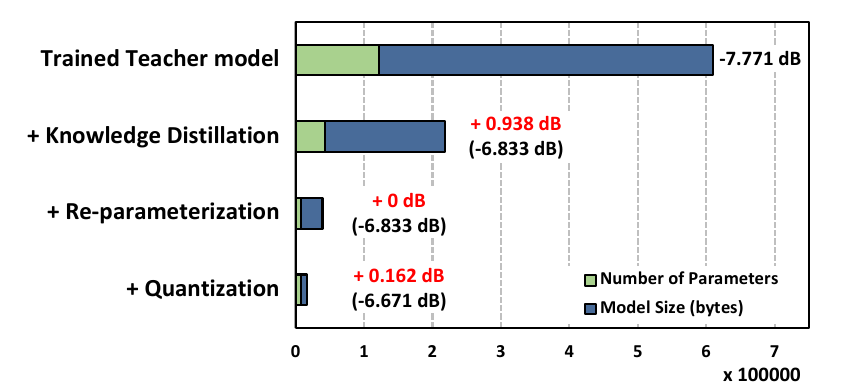}
  \vspace{-15pt}
  \caption{Effectiveness of three-stage model compression is evaluated by the parameter count, model size, and NMSE.}
  \vspace{-12pt}
  \label{fig:evaluation_compression}
\end{figure}

To evaluate the effectiveness of the multi-stage model compression strategies applied to \sysname, we analyze their progressive impact on parameter count, model size, and performance degradation, as illustrated in Fig.~\ref{fig:evaluation_compression}.
The comparison spans the teacher model and compressed models derived through three sequential steps: knowledge distillation, convolutional re-parameterization, and quantization-aware training.

The teacher model establishes the baseline, characterized by its large parameter count and correspondingly high computational demands, with \(121.9\,\mathrm{K}\) parameters and \(419.5\,\mathrm{M}\) FLOPs.
Knowledge distillation alone reduces parameters to \(35.8\%\) of the original while retaining critical channel-specific features in the student network. This initial compression incurs a modest NMSE degradation of \(0.938\,\mathrm{dB}\).
Subsequent re-parameterization further streamlines redundancy by restructuring the network’s convolutional layers, which slashes parameters by another \(80\%\) without requiring retraining.
At the third stage, quantization-aware training introduces $8$-bit integer precision for weights.
While per-channel quantization slightly increases parameter count due to the introduction of scale factors and zero-points, it further shrinks the model size to roughly a quarter compared with the previous step, with negligible NMSE degradation of \(0.162\,\mathrm{dB}\).
These results confirm the complementary efficacy of the three-stage pipeline, which ensures hardware friendliness.

\subsubsection{Generalization to Unseen Channel Models}
To evaluate the generalization capabilities of \sysname, we tested it and several baseline models on a set of unseen channel profiles. As detailed in our experimental setup, all models were trained exclusively on the CDL-B profile, which represents a complex NLoS environment. The comparative performance is presented in Table~\ref{tab:eval_unseen}.

The results reveal several key insights into the models' robustness. Our proposed algorithm demonstrates state-of-the-art performance on the unseen LoS profiles, CDL-D and CDL-E, outperforming all baselines by a substantial margin. More notably, it also achieves the best performance on the complex NLoS profile, CDL-C, indicating its strong ability to adapt to different multipath-rich environments. Conversely, a noticeable performance degradation is observed for most models, including our own, on the CDL-A profile. This is a critical finding: CDL-A, while also NLoS, represents a sparse channel with very low delay and angular spreads. The significant statistical mismatch between the complex training environment (CDL-B) and the sparse testing environment (CDL-A) poses a challenge for models that are tuned to extract complex features.

\begin{table}[t!]
\centering
\caption{Generalization performance (NMSE in -dB) on unseen channel models. \textbf{Bold} denotes the best result for each channel model.}
\label{tab:eval_unseen}
\renewcommand{\arraystretch}{1.1}
\resizebox{\linewidth}{!}{%
\begin{tabular}{l|ccccc}
\toprule
\textbf{Model} & \textbf{CDL-B} & \textbf{CDL-A} & \textbf{CDL-C} & \textbf{CDL-D} & \textbf{CDL-E} \\
& (NLoS, Seen) & (NLoS) & (NLoS) & (LoS) & (LoS) \\
\midrule
\textbf{\textit{SwiftChannel} (ours)} & 6.671 & -0.505 & \textbf{7.557} & \textbf{10.445} & \textbf{10.559} \\
\midrule
Channelformer~\cite{luan2023channelformer} & \textbf{7.600} & -0.469 & 5.423 & 9.387 & 9.024 \\
ReEsNet~\cite{li2019deep} & 5.717 & -0.977 & 6.067 & 8.685 & 8.045 \\
LSiDNN~\cite{sharma2024low} & 4.690 & \textbf{-0.259} & 2.749 & 6.644 & 6.658 \\
I-ResNet~\cite{luan2021low} & 3.633 & -1.263 & 4.310 & 6.224 & 5.534 \\
\bottomrule
\end{tabular}
}
\vspace{-8pt}
\end{table}

\subsection{Evaluation on Hardware Implementation}
This section presents a comprehensive evaluation of the proposed fine-grained pipelined hardware accelerator's performance, resource utilization, and energy efficiency through post-synthesis and post-implementation analyses.

\begin{table*}[t]
\centering
\caption{Comparison between proposed FPGA-based hardware implementation and commercial GPUs.}
\label{tab:gpu_comparison}
\vspace{-5pt}
\renewcommand{\arraystretch}{1.1}
\begin{tabular}{l|c|w{c}{1.9cm}|w{c}{1.9cm}|w{c}{1.9cm}|w{c}{1.9cm}}
\toprule
\textbf{Metric} & 
\textbf{Proposed Accelerator} & 
\multicolumn{2}{c|}{\textbf{NVIDIA Jetson Orin}} & 
\multicolumn{2}{c}{\textbf{NVIDIA RTX 3090}} \\
\midrule
Implemented & 
\makecell{Least Squares Estimator +\\Neural Network} & 
\multicolumn{2}{c|}{Neural Network} & 
\multicolumn{2}{c}{Neural Network} \\
\midrule
Latency (ms) & 
\textbf{0.883} & 
21.87 & {\(\textbf{0.04} \times\)} & 
1.830 & {\(\textbf{0.48} \times\)} \\
Throughput (FPS) & 
\textbf{1132} & 
45.72 & {\(\textbf{24.8} \times\)} & 
546.4 & {\(\textbf{2.07} \times\)} \\
Power Consumption (W) & 
\textbf{7.31} & 
11.1 & \(\textbf{0.66} \times\) & 
119 & \(\textbf{0.06} \times\) \\
Energy Efficiency (FPS/W) & 
\textbf{155} & 
4.12 & \(\textcolor{red}{\textbf{37.6} \times}\) & 
4.59 & \(\textcolor{red}{\textbf{33.8} \times}\) \\
\bottomrule
\end{tabular}
\end{table*}

\subsubsection{Physical Implementation}
While the hardware design is platform-independent, the accelerator was deployed on the Avnet ADRS1000 development board for evaluation, which features a Xilinx Zynq UltraScale+ RFSoC XCZU49DR. This heterogeneous platform combines a quad-core ARM Cortex-A53 processing system with an FPGA programmable logic fabric, enabling flexible hardware-software co-design. The design leveraged Vitis HLS as the development platform to generate optimized RTL implementations. Vivado was employed for system integration, connecting the custom IP core, DMA controllers, and the PS via AXI interconnects. Xilinx Vitis facilitated host-device communication for data transfer management.
The architecture emphasizes modularity, with distinct functional blocks for convolutional processing, attention mechanisms, and pixel shuffling operations.

\begin{figure}[t]
    \centering
    \centering
    \includegraphics[width=0.95\linewidth]{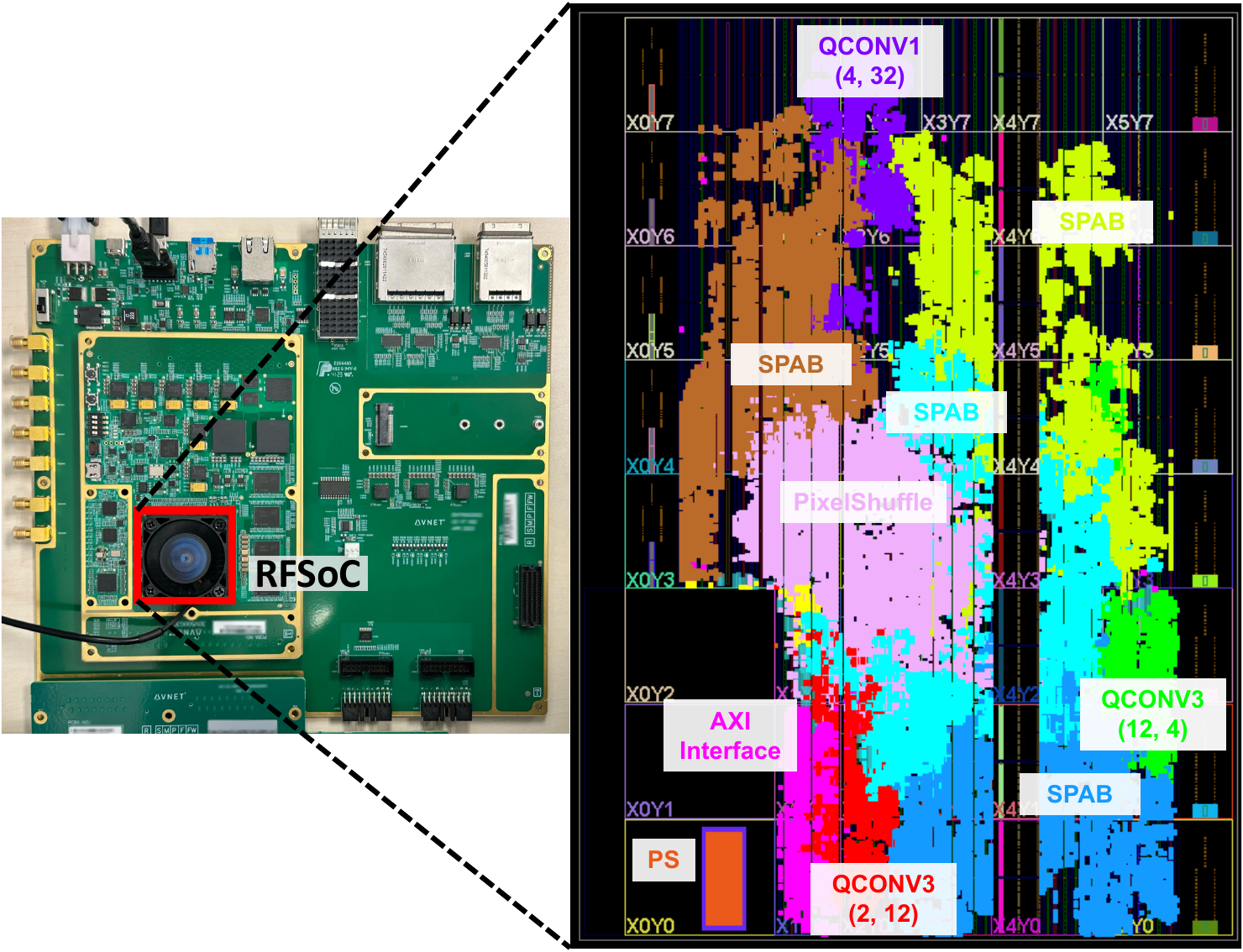}
    \caption{Board setup and physical implementation of design.}
    \label{fig:evaluation_phy_impl}
    \vspace{-12pt}
\end{figure}

Fig.~\ref{fig:evaluation_phy_impl} illustrates the physical layout of the accelerator on the RFSoC. The SPAB blocks, each integrating two cascaded QCONV3 engines and the parameter-free attention module, dominate the PL area due to their computational intensity.
Another major contributor to hardware utilization is the PixelShuffle block, which, despite requiring no arithmetic operations, demands substantial logic resources for control signal routing and dataflow orchestration.

\subsubsection{Comparison with Commercial GPUs}
Table~\ref{tab:gpu_comparison} quantifies the power performance disparity between the proposed FPGA-based accelerator and two high-performance commercial GPUs: the NVIDIA Jetson AGX ORIN and the NVIDIA RTX 3090.
Notably, the comparison excludes GPU implementations of LS estimation, as such tasks typically require specific kernel accelerations to achieve low latency. Throughput metrics are derived by inverting single-input latency measurements, given the absence of parallelized multi-input execution scenarios in this evaluation.
The proposed hardware acceleration demonstrates exceptional performance, achieving the lowest latency of \(0.883\,\mathrm{ms}\) and significantly outperforming both the Jetson ORIN and the RTX 3090 GPUs, delivering \(24.8 \times\) and \(2.07 \times\) speedup, respectively.
Energy efficiency, defined as throughput per watt, further widens this gap, with the accelerator surpassing both GPU platforms by more than \(33\) times. Such efficiency aligns well with the stringent power budgets of edge devices.

This latency-energy advantage stems from fundamental architectural differences. While GPUs excel at dense, regular workloads through massive thread-level parallelism, the proposed design leverages spatial computation and pipelined dataflow embedded in the FPGA platform to minimize idle logic and maximize utilization of DSP slices and BRAMs.
Additionally, the FPGA’s ability to employ customized precision, such as 8-bit integer computations, reduces switching activity and register pressure compared to the 32-bit floating-point dominance typical of GPUs.
Overall, the reconfigurability of FPGAs enables adaptive resource allocation for evolving wireless standards, resulting in a critical advantage in heterogeneous network deployments.

\begin{figure}[t]
  \centering
  \includegraphics[width=0.95\linewidth]{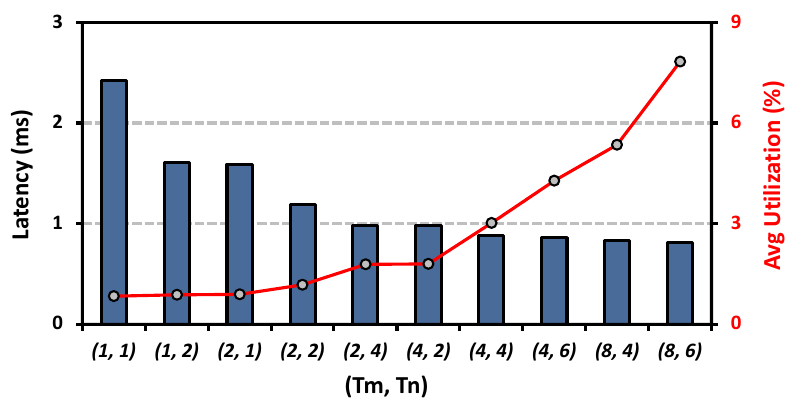}
  \vspace{-7pt}
  \caption{Design space exploration of Filter3D with 12 input channels and 8 output channels under different \(T_m\) and \(T_n\). Average utilization includes BRAMs, DSPs, FFs, and LUTs.}
  \label{fig:evaluation_dse}
  \vspace{-10pt}
\end{figure}

\begin{table*}[t]
\centering
\caption{Post-synthesis block-wise analysis. \((x, y)\) = (number of input channels, number of output channels).}
\label{tab:synthesis}
\renewcommand{\arraystretch}{0.8}
\begin{tabular}{l|c|c|c|c|c}
\toprule
\textbf{Block Name} & \textbf{Clock Cycles} & \textbf{BRAMs} & \textbf{DSPs} & \textbf{FFs} & \textbf{LUTs} \\ \midrule
\raisebox{-0.5\height}{Filter3D$_{(12, 8)}$}
& \raisebox{-0.5\height}{\colorbarClock{1}{176,473}}
& \raisebox{-0.5\height}{\colorbarBRAM{1}{73}}
& \raisebox{-0.5\height}{\colorbarDSP{1}{119}}
& \raisebox{-0.5\height}{\colorbarFF{0.221}{5,702}}
& \raisebox{-0.5\height}{\colorbarLUT{0.272}{7,991}} \\ \hline

\raisebox{-0.5\height}{QCONV1$_{(4, 32)}$}
& \raisebox{-0.5\height}{\colorbarClock{0.902}{159,193}}
& \raisebox{-0.5\height}{\colorbarBRAM{0.014}{1}}
& \raisebox{-0.5\height}{\colorbarDSP{0.664}{79}}
& \raisebox{-0.5\height}{\colorbarFF{0.437}{11,255}}
& \raisebox{-0.5\height}{\colorbarLUT{0.456}{13,429}} \\ \hline

\raisebox{-0.5\height}{PixelShuffle}
& \raisebox{-0.5\height}{\colorbarClock{0.634}{111,819}}
& \raisebox{-0.5\height}{\colorbarBRAM{0.027}{2}}
& \raisebox{-0.5\height}{\colorbarDSP{0.0}{0}}
& \raisebox{-0.5\height}{\colorbarFF{1}{25,744}}
& \raisebox{-0.5\height}{\colorbarLUT{1}{29,422}} \\ \hline

\raisebox{-0.5\height}{Output De-quantization}
& \raisebox{-0.5\height}{\colorbarClock{0.627}{110,595}}
& \raisebox{-0.5\height}{\colorbarBRAM{0.0}{0}}
& \raisebox{-0.5\height}{\colorbarDSP{0.008}{1}}
& \raisebox{-0.5\height}{\colorbarFF{0.002}{60}}
& \raisebox{-0.5\height}{\colorbarLUT{0.005}{145}} \\ \hline

\raisebox{-0.5\height}{Window3D$_{(12, 8)}$}
& \raisebox{-0.5\height}{\colorbarClock{0.237}{41,871}}
& \raisebox{-0.5\height}{\colorbarBRAM{0.027}{2}}
& \raisebox{-0.5\height}{\colorbarDSP{0.0}{0}}
& \raisebox{-0.5\height}{\colorbarFF{0.033}{853}}
& \raisebox{-0.5\height}{\colorbarLUT{0.057}{1,666}} \\ \hline

\raisebox{-0.5\height}{Attention Calculation}
& \raisebox{-0.5\height}{\colorbarClock{0.235}{41,488}}
& \raisebox{-0.5\height}{\colorbarBRAM{0.014}{1}}
& \raisebox{-0.5\height}{\colorbarDSP{0.336}{40}}
& \raisebox{-0.5\height}{\colorbarFF{0.065}{1,662}}
& \raisebox{-0.5\height}{\colorbarLUT{0.057}{1,691}} \\ \hline

\raisebox{-0.5\height}{ReLU Activation}
& \raisebox{-0.5\height}{\colorbarClock{0.157}{27,650}}
& \raisebox{-0.5\height}{\colorbarBRAM{0.0}{0}}
& \raisebox{-0.5\height}{\colorbarDSP{0.0}{0}}
& \raisebox{-0.5\height}{\colorbarFF{0.001}{29}}
& \raisebox{-0.5\height}{\colorbarLUT{0.005}{148}} \\ \hline

\raisebox{-0.5\height}{Input Quantization}
& \raisebox{-0.5\height}{\colorbarClock{0.039}{6,918}}
& \raisebox{-0.5\height}{\colorbarBRAM{0.0}{0}}
& \raisebox{-0.5\height}{\colorbarDSP{0.084}{10}}
& \raisebox{-0.5\height}{\colorbarFF{0.022}{573}}
& \raisebox{-0.5\height}{\colorbarLUT{0.022}{636}} \\ \hline

\raisebox{-0.5\height}{LS Estimator}
& \raisebox{-0.5\height}{\colorbarClock{0.039}{6,917}}
& \raisebox{-0.5\height}{\colorbarBRAM{0.014}{1}}
& \raisebox{-0.5\height}{\colorbarDSP{0.034}{4}}
& \raisebox{-0.5\height}{\colorbarFF{0.020}{506}}
& \raisebox{-0.5\height}{\colorbarLUT{0.018}{539}} \\
\bottomrule
\end{tabular}
\vspace{-5pt}
\end{table*}

\subsubsection{Design Space Exploration of QCONV3 Engine}
The fine-grained pipelined architecture of the entire design dictates that total timing is governed by its most latency-critical component, which appears to be the QCONV3 engine within the SPAB block implementing tiled convolution.
These engines are the most computationally intensive and consist of two symmetric configurations: four layers with \(F_{in}=12, F_{out}=8\) and four with \(F_{in}=8, F_{out}=12\).
The number of PEs is determined by the selection of \(T_m\) (output channels per PE) and \(T_n\) (input channels per PE), which in turn influences the whole hardware design.
To evaluate the impact of different combinations of \(T_m\) and \(T_n\), we conduct design space exploration for the Filter3D module within the engine.

To determine the optimal tiling parameters for these critical blocks, we conducted a design space exploration. Due to our tiling methodology, the resource and latency trade-offs are identical for both configurations, making the DSE performed on the representative \((F_{in}=12, F_{out}=8)\) case applicable to all eight engines.
Fig.~\ref{fig:evaluation_dse} illustrates the relationship between latency and average resource utilization for various configurations.
It is evident that selecting values for both \(T_m\) and \(T_n\) greater than \(4\) is necessary to achieve a delay of less than \(1\,\mathrm{ms}\).
However, larger values also result in an increase in hardware complexity, as each PE is required to handle more input data, contributing to a greater number of output channels.
Consequently, for our final design, we selected both \(T_m\) and \(T_n\) to be \(4\), striking a balance between latency and utilization.
The other, shallower QCONV3 layers in the design, such as those with \((F_{in}=2, F_{out}=12)\) and \((F_{in}=12, F_{out}=4)\), are not on the critical path for latency and thus did not require a separate DSE; for these layers, we applied robust tiling parameters (\(T_m=4, T_n=2\) and \(T_m=4, T_n=4\) respectively) as a safe and effective design choice.

\subsubsection{Post-Synthesis Analysis}
The synthesis phase transforms high-level hardware descriptions into gate-level net-lists composed of fundamental primitives such as block RAMs (BRAMs), digital signal processing slices (DSPs), look-up tables (LUTs), and flip-flops (FFs).
This stage provides critical insights into hardware constraints and optimization effectiveness prior to physical implementation, offering quantitative measures of algorithmic complexity and resource allocation.
As detailed in Table~\ref{tab:synthesis}, clock cycles and utilization metrics for individual blocks highlight distinct architectural challenges.
The Filter3D block within the QCONV3 engine emerges as the most resource-demanding component, consuming \(119\) DSP slices and \(73\) BRAMs due to intensive multiply-accumulate operations and bias accumulation processes, emerging as the primary bottleneck in overall timing.
Conversely, the PixelShuffle block exhibits a latency reduction from \(222\,\mathrm{K}\) to \(112\,\mathrm{K}\) clock cycles after implementing a data pipelining mechanism.
This performance gain, however, results in a necessary increase in FF and LUT utilization. This is a direct consequence of replacing a simple, large BRAM-based buffer with the more complex control logic required to manage the distributed, parallel data streams in our streaming architecture.
In contrast, auxiliary components such as quantization and de-quantization modules collectively occupy fewer logic resources, demonstrating a negligible impact on overall timing and area characteristics.

\subsubsection{Post-Implementation Analysis}
\begin{table}[t]
\centering
\caption{Hardware design post-implementation report.}
\label{tab:implementation}
\renewcommand{\arraystretch}{1.2}
\begin{tabular}{l|l|c|c}
\toprule
\textbf{Category} & \textbf{Metric} & \textbf{Value} & \textbf{Utilized (\%)} \\ \midrule

\multirow{2}{*}{\textbf{Timing}} 
& Clock Frequency (MHz) & 200 & -- \\ 
& Latency (ms) & 0.883 & -- \\ \hline

\multirow{5}{*}{\textbf{Resources}}
& LUT & 140,891 & 33.13 \\ 
& LUTRAM & 3,197 & 1.50 \\ 
& FF & 103,841 & 12.21 \\ 
& BRAM & 413.5 & 38.29 \\ 
& DSP & 1079 & 25.26 \\ \hline

\textbf{Power} & On-chip Power (W) & 7.311 & -- \\ \hline

\textbf{Accuracy} & NMSE (-dB) & 6.628 & -- \\ 
\bottomrule
\end{tabular}
\vspace{-8pt}
\end{table}
The post-implementation evaluation validates the design’s operational viability after place-and-route optimization on the target FPGA device, revealing key performance and efficiency metrics.
As summarized in Table~\ref{tab:implementation}, the proposed fine-grained pipeline architecture achieves a balanced trade-off between performance and hardware complexity, operating at a \(200\,\mathrm{MHz}\) clock frequency while delivering an end-to-end latency of \(0.883\,\mathrm{ms}\), lower than \(1\,\mathrm{ms}\) feedback deadline mandated by 5G NR standards.
Resource utilization analysis demonstrates the allocation of programmable logic, with \(33.1\%\) LUTs and \(12.2\%\) FFs dedicated to control state machines and dataflow orchestration.
Computationally intensive modules account for \(25.3\%\) DSP slice utilization in total, aligning with synthesis-stage projections.
The memory footprint remains well-optimized at \(38.29\%\) BRAM occupancy, sufficient to store all on-chip weights, biases, and intermediate feature maps without external memory access penalties. This conservative resource footprint preserves the remaining logic resources for concurrent execution of other protocol-specific tasks, underscoring the architecture’s adaptability to multi-functional wireless systems.
Power analysis indicates a total board-level consumption of \(7.311\,\mathrm{W}\), underscoring the energy efficiency of the design.
Functional verification against a comprehensive dataset confirms that the bitstream accuracy matches software simulation outputs, validating the performance of the deployed accelerator.

\section{Conclusion}
\label{sec:conclusion}
In this work, we propose \sysname, a novel algorithm-hardware co-design framework for 5G MIMO channel estimation that achieves high-accuracy channel reconstruction with sub-millisecond latency.
The proposed deep learning model incorporates a parameter-free attention mechanism to efficiently reconstruct channel matrices from LS estimates in the spatial-frequency domain.
Advanced multi-stage model compression techniques substantially reduce computational complexity and memory footprint while preserving estimation accuracy, yielding a hardware-efficient architecture.
A dedicated accelerator targeting the FPGA platform is specially designed for the proposed algorithm, demonstrating superior latency and power efficiency compared to commercial GPU implementations.
Extensive experiments validate the framework’s robustness, demonstrating strong generalization not only across diverse SNR and mobility scenarios but also, critically, to a variety of unseen channel models, which confirms its applicability to real-world 5G systems.

\section*{Acknowledgment}
This work was supported by the NSF of Guangdong Province (Project No. 2024A1515010192), Research Grants Council of the Hong Kong Special Administrative Region, China (Project No. CityU 11202925 and CityU 11202124), the Innovation and Technology Commission of Hong Kong (Project No. MHP/072/23), and the CityU SRG-Fd (Project No. 11205323).
This work was conducted when the first author visited the National Engineering Laboratory for Big Data System Computing Technology at Shenzhen University.

\ifCLASSOPTIONcaptionsoff
  \newpage
\fi

\bibliographystyle{IEEEtran}
\bibliography{citations}

@article{larsson2014massive,
  title={Massive MIMO for next generation wireless systems},
  author={Larsson, Erik G and Edfors, Ove and Tufvesson, Fredrik and Marzetta, Thomas L},
  journal={IEEE Communications Magazine},
  volume={52},
  number={2},
  pages={186--195},
  year={2014}
}

@misc{3gpp.38.211,
  author = "{3GPP}",
  title = "{5G; NR; Physical Channels and Modulation}",
  institution = "3rd Generation Partnership Project (3GPP)",
  year = 2024,
  month = oct,
  note = "3GPP TS 38.211 version 18.4.0 Release 18",
}

@inproceedings{wan2024swift,
  title={Swift parameter-free attention network for efficient super-resolution},
  author={Wan, Cheng and Yu, Hongyuan and Li, Zhiqi and Chen, Yihang and Zou, Yajun and Liu, Yuqing and Yin, Xuanwu and Zuo, Kunlong},
  booktitle={Proceedings of the IEEE/CVF Conference on Computer Vision and Pattern Recognition (CVPR)},
  pages={6246--6256},
  year={2024}
}

@inproceedings{choe2019attention,
  title={Attention-based dropout layer for weakly supervised object localization},
  author={Choe, Junsuk and Shim, Hyunjung},
  booktitle={Proceedings of the IEEE/CVF Conference on Computer Vision and Pattern Recognition (CVPR)},
  pages={2219--2228},
  year={2019}
}

@article{du2022parameter,
  title={Parameter-free similarity-aware attention module for medical image classification and segmentation},
  author={Du, Jie and Guan, Kai and Zhou, Yanhong and Li, Yuanman and Wang, Tianfu},
  journal={IEEE Transactions on Emerging Topics in Computational Intelligence (TETCI)},
  volume={7},
  number={3},
  pages={845--857},
  year={2022}
}

@article{shi2023parameter,
  title={Parameter-free channel attention for image classification and super-resolution},
  author={Shi, Yuxuan and Yang, Lingxiao and An, Wangpeng and Zhen, Xiantong and Wang, Liuqing},
  journal={arXiv preprint arXiv:2303.11055 (arXiv)},
  year={2023}
}

@inproceedings{yang2021simam,
  title={Simam: A simple, parameter-free attention module for convolutional neural networks},
  author={Yang, Lingxiao and Zhang, Ru-Yuan and Li, Lida and Xie, Xiaohua},
  booktitle={Proceedings of the 38th International Conference on Machine Learning (ICML)},
  pages={11863--11874},
  year={2021}
}

@article{vaswani2017attention,
  title={Attention is all you need},
  author={Vaswani, Ashish and Shazeer, Noam and Parmar, Niki and Uszkoreit, Jakob and Jones, Llion and Gomez, Aidan N and Kaiser, {\L}ukasz and Polosukhin, Illia},
  journal={Advances in neural information processing systems},
  volume={30},
  year={2017}
}

@inproceedings{ding2021repvgg,
  title={Repvgg: Making vgg-style convnets great again},
  author={Ding, Xiaohan and Zhang, Xiangyu and Ma, Ningning and Han, Jungong and Ding, Guiguang and Sun, Jian},
  booktitle={Proceedings of the IEEE/CVF Conference on Computer Vision and Pattern Recognition (CVPR)},
  pages={13733--13742},
  year={2021}
}

@inproceedings{van1995channel,
  title={On channel estimation in OFDM systems},
  author={Van De Beek, J-J and Edfors, Ove and Sandell, Magnus and Wilson, Sarah Kate and Borjesson, P Ola},
  booktitle={Proceedings of the 45th Vehicular Technology Conference (VTC)},
  volume={2},
  pages={815--819},
  year={1995}
}

@article{li1998robust,
  title={Robust channel estimation for OFDM systems with rapid dispersive fading channels},
  author={Li, YLJC and Cimini, Leonard J and Sollenberger, Nelson R},
  journal={IEEE Transactions on Communications (TCOM)},
  volume={46},
  number={7},
  pages={902--915},
  year={1998}
}

@article{alwazani2020intelligent,
  title={Intelligent reflecting surface-assisted multi-user MISO communication: Channel estimation and beamforming design},
  author={Alwazani, Hibatallah and Kammoun, Abla and Chaaban, Anas and Debbah, M{\'e}rouane and Alouini, Mohamed-Slim and others},
  journal={IEEE Open Journal of the Communications Society},
  volume={1},
  pages={661--680},
  year={2020}
}

@article{pourkabirian2021robust,
  title={Robust channel estimation in multiuser downlink 5G systems under channel uncertainties},
  author={Pourkabirian, Azadeh and Anisi, Mohammad Hossein},
  journal={IEEE Transactions on Mobile Computing (TMC)},
  volume={21},
  number={12},
  pages={4569--4582},
  year={2021}
}

@inproceedings{dong2016accelerating,
  title={Accelerating the super-resolution convolutional neural network},
  author={Dong, Chao and Loy, Chen Change and Tang, Xiaoou},
  booktitle={Proceedings of European Conference on Computer Vision (ECCV)},
  pages={391--407},
  year={2016}
}

@article{soltani2019deep,
  title={Deep learning-based channel estimation},
  author={Soltani, Mehran and Pourahmadi, Vahid and Mirzaei, Ali and Sheikhzadeh, Hamid},
  journal={IEEE Communications Letters (COMML)},
  volume={23},
  number={4},
  pages={652--655},
  year={2019}
}

@article{li2019deep,
  title={Deep residual learning meets OFDM channel estimation},
  author={Li, Lianjun and Chen, Hao and Chang, Hao-Hsuan and Liu, Lingjia},
  journal={IEEE Wireless Communications Letters (WCL)},
  volume={9},
  number={5},
  pages={615--618},
  year={2019}
}

@inproceedings{luan2021low,
  title={Low complexity channel estimation with neural network solutions},
  author={Luan, Dianxin and Thompson, John},
  booktitle={Proceedings of the 25th International ITG Workshop on Smart Antennas (WSA)},
  pages={1--6},
  year={2021}
}

@article{luan2023channelformer,
  title={Channelformer: Attention based neural solution for wireless channel estimation and effective online training},
  author={Luan, Dianxin and Thompson, John S},
  journal={IEEE Transactions on Wireless Communications (TWC)},
  volume={22},
  number={10},
  pages={6562--6577},
  year={2023}
}

@inproceedings{yang2021deep,
  title={Deep learning based OFDM channel estimation using frequency-time division and attention mechanism},
  author={Yang, Ang and Sun, Peng and Rakesh, Tamrakar and Sun, Bule and Qin, Fei},
  booktitle={Proceedings of the IEEE Globecom Workshops},
  pages={1--6},
  year={2021}
}

@article{zhou2023pay,
  title={Pay less but get more: A dual-attention-based channel estimation network for massive MIMO systems with low-density pilots},
  author={Zhou, Binggui and Yang, Xi and Ma, Shaodan and Gao, Feifei and Yang, Guanghua},
  journal={IEEE Transactions on Wireless Communications (TWC)},
  pages={6061--6076},
  year={2023}
}

@article{zhou2025low,
  title={Low-Overhead Channel Estimation via 3D Extrapolation for TDD mmWave Massive MIMO Systems Under High-Mobility Scenarios},
  author={Zhou, Binggui and Yang, Xi and Ma, Shaodan and Gao, Feifei and Yang, Guanghua},
  journal={IEEE Transactions on Wireless Communications (TWC)},
  year={2025}
}

@article{sharma2024low,
  title={Low complexity deep learning augmented wireless channel estimation for pilot-based ofdm on zynq system on chip},
  author={Sharma, Animesh and Haq, Syed Asrar Ul and Darak, Sumit J},
  journal={IEEE Transactions on Circuits and Systems I: Regular Papers (TCAS-I)},
  year={2024}
}

@article{lin2022channel,
  title={Channel estimation for IRS-assisted millimeter-wave MIMO systems: Sparsity-inspired approaches},
  author={Lin, Tian and Yu, Xianghao and Zhu, Yu and Schober, Robert},
  journal={IEEE Transactions on Communications (TCOM)},
  volume={70},
  number={6},
  pages={4078--4092},
  year={2022}
}

@article{chen2023channel,
  title={Channel estimation for reconfigurable intelligent surface aided multi-user mmWave MIMO systems},
  author={Chen, Jie and Liang, Ying-Chang and Cheng, Hei Victor and Yu, Wei},
  journal={IEEE Transactions on Wireless Communications (TWC)},
  volume={22},
  number={10},
  pages={6853--6869},
  year={2023}
}

@article{gao2015spatially,
  title={Spatially common sparsity based adaptive channel estimation and feedback for FDD massive MIMO},
  author={Gao, Zhen and Dai, Linglong and Wang, Zhaocheng and Chen, Sheng},
  journal={IEEE Transactions on Signal Processing (TSP)},
  volume={63},
  number={23},
  pages={6169--6183},
  year={2015}
}

@article{jiang2021dual,
  title={Dual CNN-based channel estimation for MIMO-OFDM systems},
  author={Jiang, Peiwen and Wen, Chao-Kai and Jin, Shi and Li, Geoffrey Ye},
  journal={IEEE Transactions on Communications (TCOM)},
  volume={69},
  number={9},
  pages={5859--5872},
  year={2021}
}

@article{dong2019deep,
  title={Deep CNN-based channel estimation for mmWave massive MIMO systems},
  author={Dong, Peihao and Zhang, Hua and Li, Geoffrey Ye and Gaspar, Ivan Simoes and NaderiAlizadeh, Navid},
  journal={IEEE Journal of Selected Topics in Signal Processing (JSTSP)},
  volume={13},
  number={5},
  pages={989--1000},
  year={2019}
}

@article{jiang2021learning,
  title={Learning to reflect and to beamform for intelligent reflecting surface with implicit channel estimation},
  author={Jiang, Tao and Cheng, Hei Victor and Yu, Wei},
  journal={IEEE Journal on Selected Areas in Communications (JSAC)},
  volume={39},
  number={7},
  pages={1931--1945},
  year={2021}
}

@article{chundi2021channel,
  title={Channel estimation using deep learning on an FPGA for 5G millimeter-wave communication systems},
  author={Chundi, Pavan Kumar and Wang, Xiaodong and Seok, Mingoo},
  journal={IEEE Transactions on Circuits and Systems I: Regular Papers (TCAS-I)},
  volume={69},
  number={2},
  pages={908--918},
  year={2021}
}

@article{mirfarshbafan2020beamspace,
  title={Beamspace channel estimation for massive MIMO mmWave systems: Algorithm and VLSI design},
  author={Mirfarshbafan, Seyed Hadi and Gallyas-Sanhueza, Alexandra and Ghods, Ramina and Studer, Christoph},
  journal={IEEE Transactions on Circuits and Systems I: Regular Papers (TCAS-I)},
  volume={67},
  number={12},
  pages={5482--5495},
  year={2020}
}

@article{haq2023deep,
  title={Deep neural network augmented wireless channel estimation for preamble-based ofdm phy on zynq system on chip},
  author={Haq, Syed Asrar Ul and Gizzini, Abdul Karim and Shrey, Shakti and Darak, Sumit J and Saurabh, Sneh and Chafii, Marwa},
  journal={IEEE Transactions on Very Large Scale Integration (VLSI) Systems},
  volume={31},
  number={7},
  pages={1026--1038},
  year={2023}
}

@inproceedings{haq2024low,
  title={Low complexity high speed deep neural network augmented wireless channel estimation},
  author={Haq, Syed Asrar Ul and Singh, Varun and Tanaji, Bhanu Teja and Darak, Sumit},
  booktitle={Proceedings of the 37th International Conference on VLSI Design and 23rd International Conference on Embedded Systems (VLSID)},
  pages={235--240},
  year={2024}
}

@article{gao2023deep,
  title={Deep learning-based channel estimation for wideband hybrid mmWave massive MIMO},
  author={Gao, Jiabao and Zhong, Caijun and Li, Geoffrey Ye and Soriaga, Joseph B and Behboodi, Arash},
  journal={IEEE Transactions on Communications (TCOM)},
  volume={71},
  number={6},
  pages={3679--3693},
  year={2023}
}

@article{gao2018comnet,
  title={ComNet: Combination of deep learning and expert knowledge in OFDM receivers},
  author={Gao, Xuanxuan and Jin, Shi and Wen, Chao-Kai and Li, Geoffrey Ye},
  journal={IEEE Communications Letters (COMML)},
  volume={22},
  number={12},
  pages={2627--2630},
  year={2018}
}

@article{jiang2022accurate,
  title={Accurate channel prediction based on transformer: Making mobility negligible},
  author={Jiang, Hao and Cui, Mingyao and Ng, Derrick Wing Kwan and Dai, Linglong},
  journal={IEEE Journal on Selected Areas in Communications (JSAC)},
  volume={40},
  number={9},
  pages={2717--2732},
  year={2022}
}

@article{hinton2015distilling,
  title={Distilling the Knowledge in a Neural Network},
  author={Hinton, Geoffrey},
  journal={arXiv preprint arXiv:1503.02531 (arXiv)},
  year={2015}
}

@article{romero2014fitnets,
  title={Fitnets: Hints for thin deep nets},
  author={Romero, Adriana and Ballas, Nicolas and Kahou, Samira Ebrahimi and Chassang, Antoine and Gatta, Carlo and Bengio, Yoshua},
  journal={arXiv preprint arXiv:1412.6550 (arXiv)},
  year={2014}
}

@article{zagoruyko2016paying,
  title={Paying more attention to attention: Improving the performance of convolutional neural networks via attention transfer},
  author={Zagoruyko, Sergey and Komodakis, Nikos},
  journal={arXiv preprint arXiv:1612.03928 (arXiv)},
  year={2016}
}

@inproceedings{tung2019similarity,
  title={Similarity-preserving knowledge distillation},
  author={Tung, Frederick and Mori, Greg},
  booktitle={Proceedings of the IEEE/CVF International Conference on Computer Vision (ICCV)},
  pages={1365--1374},
  year={2019}
}

@article{weng2024tailor,
  title={Tailor: Altering skip connections for resource-efficient inference},
  author={Weng, Olivia and Marcano, Gabriel and Loncar, Vladimir and Khodamoradi, Alireza and Sheybani, Nojan and Meza, Andres and Koushanfar, Farinaz and Denolf, Kristof and Duarte, Javier Mauricio and Kastner, Ryan},
  journal={ACM Transactions on Reconfigurable Technology and Systems (TRETS)},
  volume={17},
  number={1},
  pages={1--23},
  year={2024}
}

@article{li2020residual,
  title={Residual distillation: Towards portable deep neural networks without shortcuts},
  author={Li, Guilin and Zhang, Junlei and Wang, Yunhe and Liu, Chuanjian and Tan, Matthias and Lin, Yunfeng and Zhang, Wei and Feng, Jiashi and Zhang, Tong},
  journal={Proceedings of the 34th International Conference on Neural Information Processing Systems (NeurIPS)},
  volume={33},
  pages={8935--8946},
  year={2020}
}

@inproceedings{chang2021mix,
  title={Mix and match: A novel fpga-centric deep neural network quantization framework},
  author={Chang, Sung-En and Li, Yanyu and Sun, Mengshu and Shi, Runbin and So, Hayden K-H and Qian, Xuehai and Wang, Yanzhi and Lin, Xue},
  booktitle={Proceedings of the IEEE International Symposium on High-Performance Computer Architecture (HPCA)},
  pages={208--220},
  year={2021}
}

@article{guo2017angel,
  title={Angel-eye: A complete design flow for mapping CNN onto embedded FPGA},
  author={Guo, Kaiyuan and Sui, Lingzhi and Qiu, Jiantao and Yu, Jincheng and Wang, Junbin and Yao, Song and Han, Song and Wang, Yu and Yang, Huazhong},
  journal={IEEE Transactions on Computer-Aided Design of Integrated Circuits and Systems (TCAD)},
  volume={37},
  number={1},
  pages={35--47},
  year={2017}
}

@inproceedings{sun2022film,
  title={Film-qnn: Efficient fpga acceleration of deep neural networks with intra-layer, mixed-precision quantization},
  author={Sun, Mengshu and Li, Zhengang and Lu, Alec and Li, Yanyu and Chang, Sung-En and Ma, Xiaolong and Lin, Xue and Fang, Zhenman},
  booktitle={Proceedings of the ACM/SIGDA International Symposium on Field-Programmable Gate Arrays (ISFPGA)},
  pages={134--145},
  year={2022}
}

@article{liang2019evaluating,
  title={Evaluating fast algorithms for convolutional neural networks on FPGAs},
  author={Liang, Yun and Lu, Liqiang and Xiao, Qingcheng and Yan, Shengen},
  journal={IEEE Transactions on Computer-Aided Design of Integrated Circuits and Systems (TCAD)},
  volume={39},
  number={4},
  pages={857--870},
  year={2019}
}

@inproceedings{bai2020unified,
  title={A unified hardware architecture for convolutions and deconvolutions in CNN},
  author={Bai, Lin and Lyu, Yecheng and Huang, Xinming},
  booktitle={Proceedings of the IEEE International Symposium on Circuits and Systems (ISCAS)},
  pages={1--5},
  year={2020}
}

@inproceedings{ujjainkar2023imagen,
  title={ImaGen: A general framework for generating memory-and power-efficient image processing accelerators},
  author={Ujjainkar, Nisarg and Leng, Jingwen and Zhu, Yuhao},
  booktitle={Proceedings of the 50th Annual International Symposium on Computer Architecture (ISCA)},
  pages={1--13},
  year={2023}
}

@inproceedings{ma2017optimizing,
  title={Optimizing loop operation and dataflow in FPGA acceleration of deep convolutional neural networks},
  author={Ma, Yufei and Cao, Yu and Vrudhula, Sarma and Seo, Jae-sun},
  booktitle={Proceedings of the ACM/SIGDA International Symposium on Field-Programmable Gate Arrays (ISFPGA)},
  pages={45--54},
  year={2017}
}

@inproceedings{zhang2015optimizing,
  title={Optimizing FPGA-based accelerator design for deep convolutional neural networks},
  author={Zhang, Chen and Li, Peng and Sun, Guangyu and Guan, Yijin and Xiao, Bingjun and Cong, Jason},
  booktitle={Proceedings of the ACM/SIGDA International Symposium on Field-Programmable Gate Arrays (ISFPGA)},
  pages={161--170},
  year={2015}
}

\vspace{50pt}
\begin{IEEEbiography}[{\includegraphics[width=1in,height=1.25in,clip,keepaspectratio]{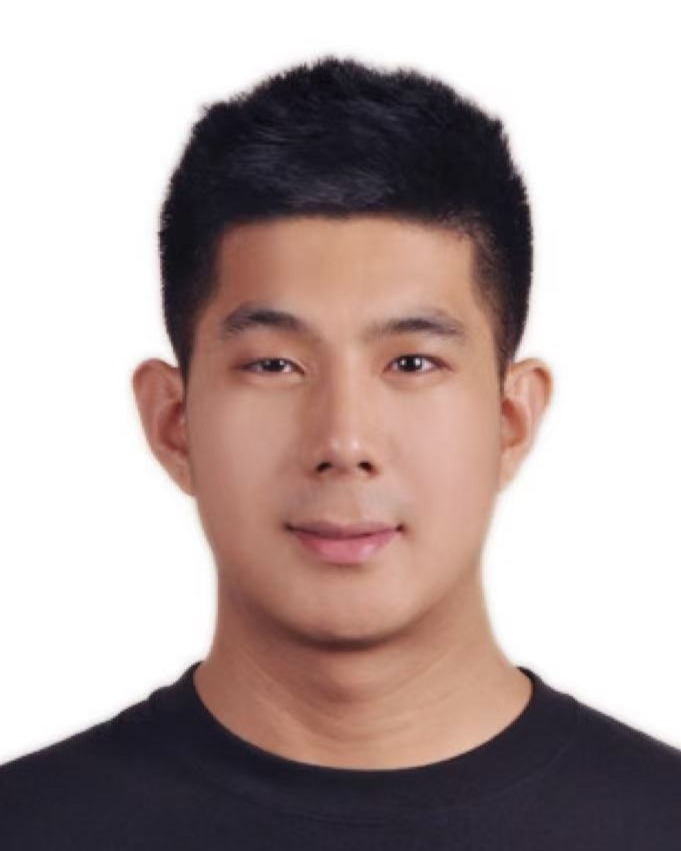}
}]{Shengzhe~Lyu}(Student Member, IEEE) is a Ph.D. candidate in the Department of Computer Science at City University of Hong Kong. Before that, he received the B.Sc. (summa cum laude) degree in electronics engineering from KU Leuven (KUL), Belgium, and the B.Eng. degree in microelectronics science and engineering from South China University of Technology (SCUT), China, both in 2023. His research focuses on software-hardware co-design and AI accelerators on FPGAs.
\end{IEEEbiography}

\begin{IEEEbiography}[{\includegraphics[width=1in,height=1.25in,clip,keepaspectratio]{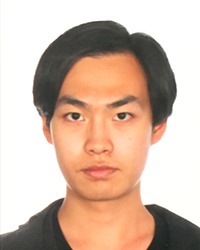}
}]{Yuhan~She}(Student Member, IEEE) received his B.Eng. degree in Electrical Engineering from Zhejiang University in 2021. He is now a Ph.D. candidate in the Department of Electrical Engineering at City University of Hong Kong. His research interests include High-Level Synthesis (HLS), Electronic Design Automation (EDA), and RISC-V.
\end{IEEEbiography}

\begin{IEEEbiography}[{\includegraphics[width=1in,height=1.25in,clip,keepaspectratio]{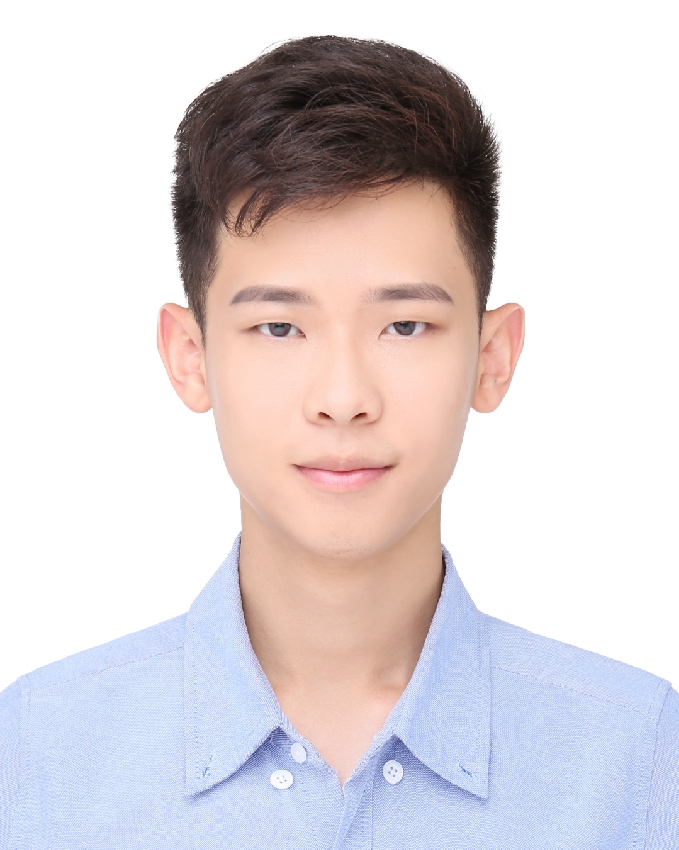}
}]{Di~Duan}(Member, IEEE), Ph.D., is currently a Postdoctoral Fellow in the Department of Information Engineering at The Chinese University of Hong Kong (CUHK). He received his Ph.D. in Computer Science from the City University of Hong Kong (CityUHK). His research lies at the intersection of mobile sensing and human-computer interaction, with a special focus on wearable and wireless sensing systems, software-hardware co-design, health and AI.
\end{IEEEbiography}

\begin{IEEEbiography}[{\includegraphics[width=1in,height=1.25in,clip,keepaspectratio]{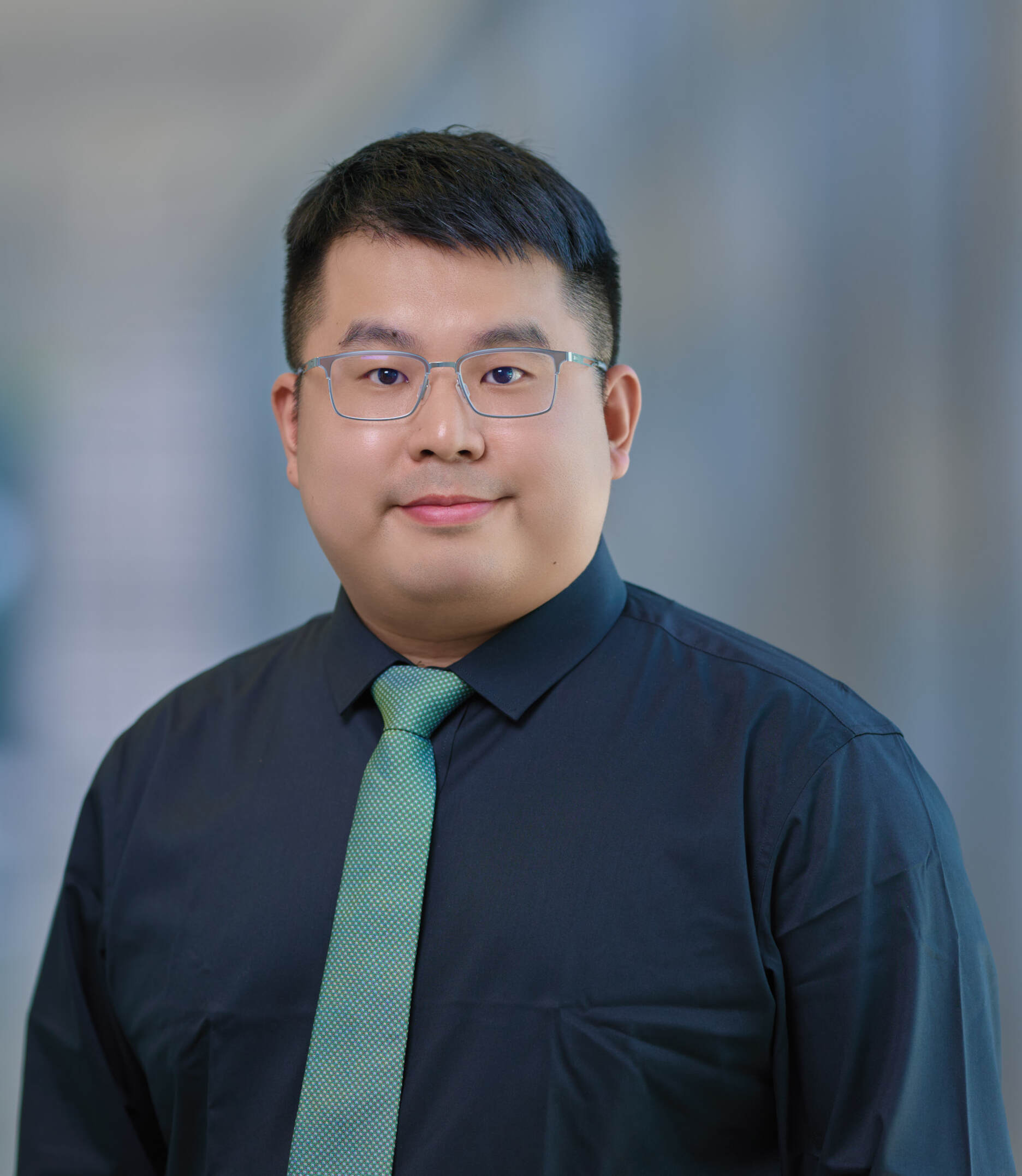}
}]{Tao Ni}(Member, IEEE) is an Assistant Professor at the Computer, Electrical and Mathematical Sciences and Engineering (CEMSE) Division, King Abdullah University of Science and Technology (KAUST).
He was a Postdoctoral Research Fellow at the Department of Computer Science, City University of Hong Kong, where he obtained his PhD. Before that, he received his Master's degree from the Australian National University in 2020 and his Bachelor's degree from Shanghai Jiao Tong University in 2018. 
His research interests include cyber-physical systems (CPS) security, AI security, and mobile computing. His work received the 2024 Cybersecurity Best Practical Paper Award and was also named an ACM MobiSys Rising Star (2024) and an AIoTSys Rising Star (2025).
\end{IEEEbiography}

\begin{IEEEbiography}
[{\includegraphics[width=1in,height=1.25in,clip,keepaspectratio]{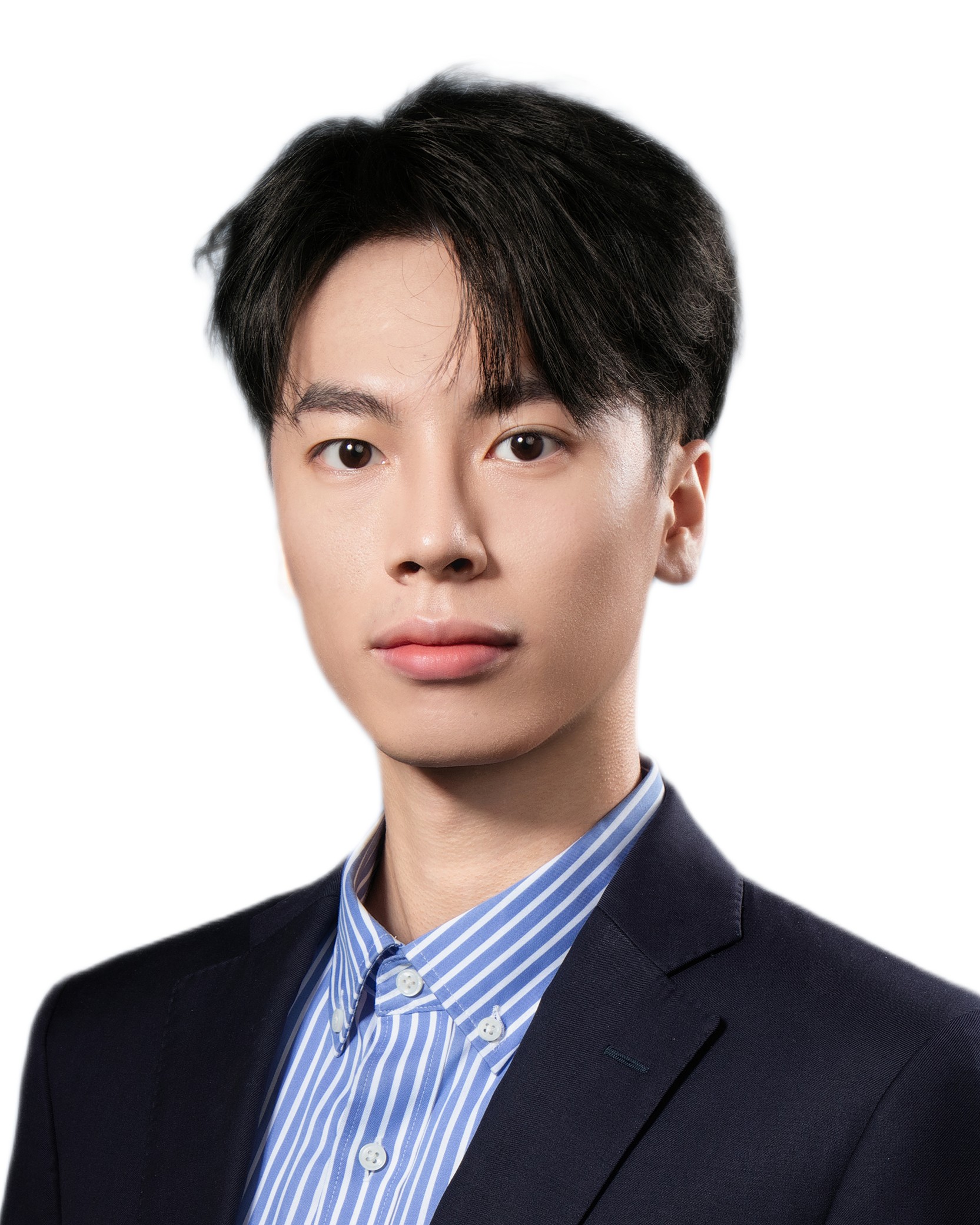}
}]{Yu~Hin~Chan}received his B.Eng. degree in Electronic and Communication Engineering from City University of Hong Kong in 2022, and his M.Sc. degree in Computing from Cardiff University, UK, in 2024. He is currently a Research Assistant at the Department of Electrical Engineering, City University of Hong Kong. His research interests include hardware-software co-design, secure communication protocols for RISC-V systems, and cryptography.
\end{IEEEbiography}

\begin{IEEEbiography}
[{\includegraphics[width=1in,height=1.25in,clip,keepaspectratio]{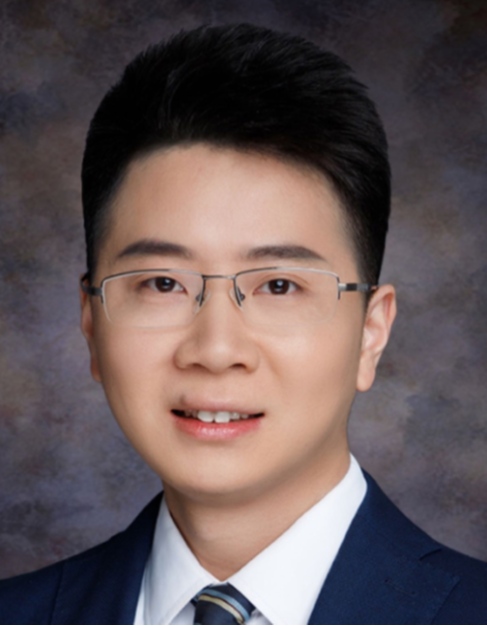}
}]{Chengwen~Luo}(Member, IEEE) received the PhD degree from the School of Computing, National University of Singapore (NUS), Singapore. He is currently a Professor with the National Engineering Laboratory for Big Data System Computing Technology, Shenzhen University (SZU), China. His research interests include mobile and pervasive computing and security aspects of Internet of Things.
\end{IEEEbiography}

\begin{IEEEbiography}[{\includegraphics[width=1in,height=1.25in,clip,keepaspectratio]{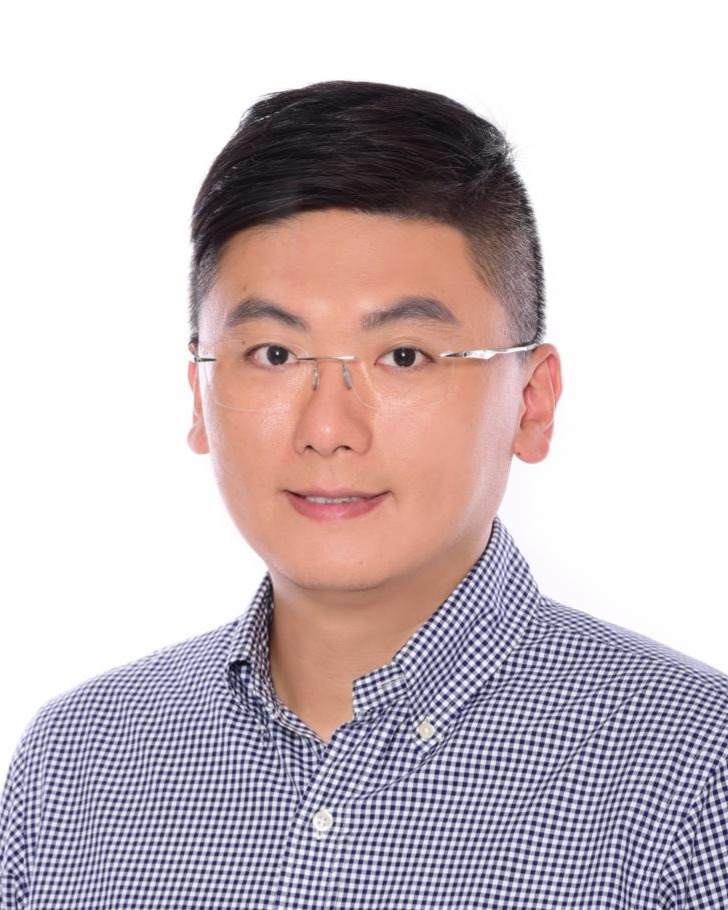}
}]{Ray~C.~C.~Cheung}(Senior Member, IEEE) received the B.Eng. (Hons.) and M.Phil. degrees in computer engineering and computer science and engineering from The Chinese University of Hong Kong (CUHK), Hong Kong, in 1999 and 2001, respectively, and the D.I.C. and Ph.D. degrees in computing from Imperial College London (IC), London, U.K., in 2007. He received the Hong Kong Croucher Foundation Fellowship for his postdoctoral research work with the Electrical Engineering Department, University of California at Los Angeles (UCLA), Los Angeles, CA, USA, and completed his visiting fellowship with Princeton University, Princeton, NJ, USA. He is a Professor with the Department of Electrical Engineering, and an Associate Provost (Digital Learning) with the City University of Hong Kong (CityUHK), Hong Kong. His current research interests include cryptographic processor designs and embedded system designs. He is currently the Chair of the IEEE Hong Kong Section.
\end{IEEEbiography}

\begin{IEEEbiography}[{\includegraphics[width=1in,height=1.25in,clip,keepaspectratio]{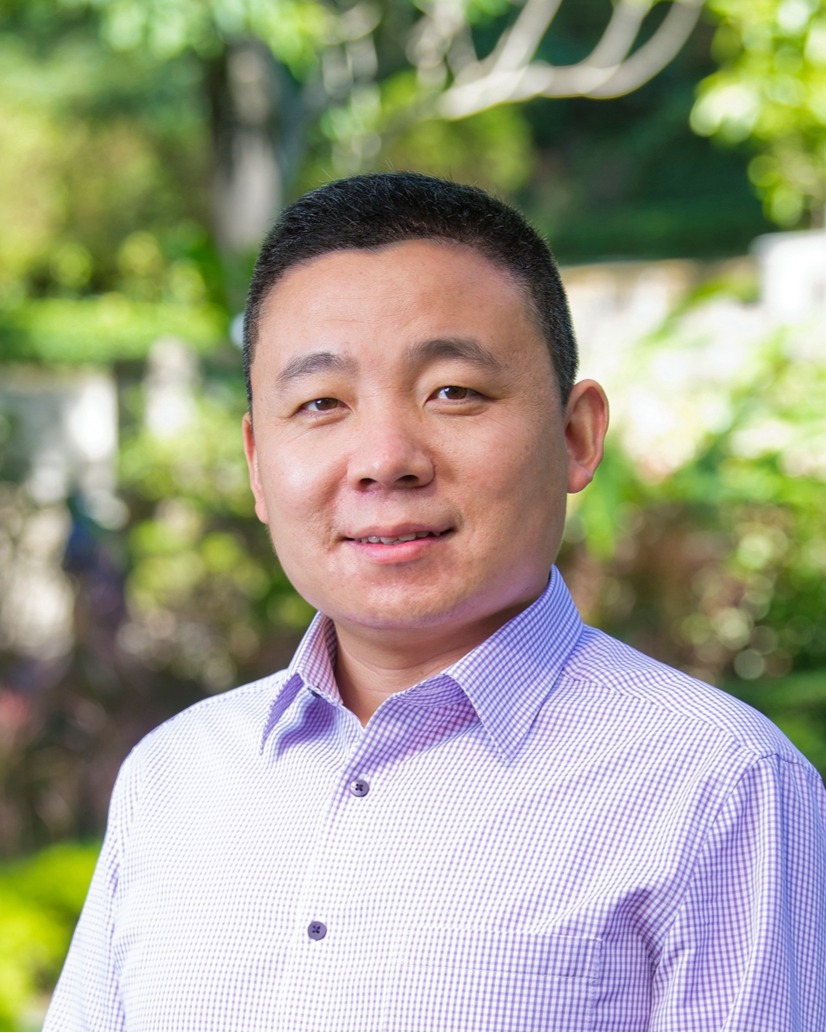}
}]{Weitao~Xu~}(Senior Member, IEEE) is an Associate Professor at the Department of Computer Science at City University of Hong Kong. He obtained his PhD degree from the University of Queensland in 2017. His research generally focuses on IoT such as smart sensing, IoT security, IoT+AI, and wireless networks. He is a senior member of IEEE.
\end{IEEEbiography}



\end{document}